\documentclass{article}

\usepackage{authblk}

\usepackage[backend=biber,
style=alphabetic,
sorting=nyt,maxbibnames=99]{biblatex}
\usepackage{amsmath,amsthm,amsfonts,amssymb}
\usepackage{mathtools}
\usepackage{physics}
\usepackage{braket}
\usepackage{array}
\allowdisplaybreaks

\usepackage[margin=2cm]{geometry}
\usepackage{xspace}
\usepackage{xcolor}
\usepackage{subcaption}

\usepackage{hyperref}
\hypersetup{
    colorlinks=true,
    linkcolor=black,
    citecolor=black,
    filecolor=black,
    urlcolor=black,
}
\usepackage[capitalize]{cleveref}
\newtheorem{theorem}{Theorem}[section]

\newtheorem{corollary}[theorem]{Corollary}

\newtheorem{question}{Question}
\theoremstyle{definition}
\newtheorem{definition}[theorem]{Definition}
\newcommand{\portnum}{port-numbering\xspace}

\newcommand{\local}{LOCAL\xspace}
\newcommand{\qlocal}{quantum-\local}
\newcommand{\detlocal}{det-\local}
\newcommand{\randlocal}{rand-\local} 
\newcommand{\slocal}{S\local}
\newcommand{\olocal}{online-\local} 

\newcommand{\dep}{dependence\xspace}
\newcommand{\dept}{dependent\xspace}
\newcommand{\boundep}{bounded-\dep}

\newcommand{\boundept}{bounded-\dept}

\newcommand{\nonsign}{non-signaling\xspace}
\newcommand{\philocal}{\(\varphi\)-\local}

\newcommand{\congest}{CONGEST\xspace}
\newcommand{\qcongest}{quantum-\congest}

\newcommand{\myO}[1]{O\!\left(#1\right)}
\newcommand{\myTheta}[1]{\Theta\!\left(#1\right)}
\newcommand{\myOmega}[1]{\Omega\!\left(#1\right)}
\newcommand{\mylittleo}[1]{o\!\left(#1\right)}
\newcommand{\mylittleomega}[1]{\omega\!\left(#1\right)}

\newcommand{\dist}{\text{dist}} %
\newcommand{\neighborhood}{\NN}
\newcommand{\indeg}{\text{indeg}} %
\newcommand{\outdeg}{\text{outdeg}} %
\newcommand{\poly}{\text{poly}} %
\newcommand{\ceil}[1]{\left \lceil #1 \right \rceil}
\newcommand{\floor}[1]{\left \lfloor #1 \right \rfloor}
\newcommand{\view}{\VV}
\newcommand{\diam}{\text{diam}}

\newcommand{\nats}{\mathbb{N}}
\newcommand{\natsPos}{\nats_{+}}

\renewcommand{\AA}{\mathcal{A}}
\newcommand{\BB}{\mathcal{B}}

\newcommand{\FF}{\mathcal{F}}

\newcommand{\NN}{\mathcal{N}}

\renewcommand{\SS}{\mathcal{S}}

\newcommand{\VV}{\mathcal{V}}

\newcommand{\st}{\ \middle| \ }
\newcommand{\alphabet}{\Sigma}
\newcommand{\inputSet}{\alphabet_{\text{in}}}
\newcommand{\outputSet}{\alphabet_{\text{out}}}
\newcommand{\inpt}{\text{in}}
\newcommand{\oupt}{\text{out}}
\newcommand{\localVar}{\mathrm{x}}
\newcommand{\problem}{\Pi}
\newcommand{\maxDeg}{\Delta}
\newcommand{\algo}{\AA}
\newcommand{\outcome}{\mathrm{O}}

\title{On the limits of distributed quantum computing}
\author[*]{Francesco d'Amore}
\affil[*]{{Bocconi University, BIDSA, Italy \protect \\ Gran Sasso Science Institute, Italy \protect \\ francesco.damore@gssi.it}}
\date{}

\begin{document}
\maketitle
\begin{abstract}
Quantum advantage is well-established in centralized computing, where quantum algorithms can solve certain problems exponentially faster than classical ones. 
In the distributed setting, significant progress has been made in bandwidth-limited networks, where quantum distributed networks have shown computational advantages over classical counterparts. 
However, the potential of quantum computing in networks that are constrained only by large distances is not yet understood.
We focus on the LOCAL model of computation (Linial, FOCS 1987), a distributed computational model where computational power and communication bandwidth are unconstrained, and its quantum generalization. 
In this brief survey, we summarize recent progress on the quantum-LOCAL model outlining its limitations with respect to its classical counterpart: we discuss emerging techniques, and highlight open research questions that could guide future efforts in the field.
\end{abstract} %
\section{Introduction}\label{sec:introduction}

Since the advent of quantum computing, extensive research has been conducted to explore its potential, revealing its advantage over classical computing, at least from a theoretical point of view: there are problems that classical algorithms solve in super-polynomial time, while quantum algorithms can solve them in polynomial time \cite{aaronson2022}.
But what about \emph{distributed} quantum advantage? 

There is a large body of research investigating this question in \emph{bandwidth-limited} networks \cite{censorhillel2022,legall2018,wu2022,wang2021,izumi2019,izumi2020,apeldoorn2022,fraigniaud2024,magniez2022}. Such networks are captured by (possibly variants of) the \congest model of computing. 
In essence, the question is: 
can a synchronous network of quantum computers that send \( b \) quantum qubits per time unit to each neighbor outperform a synchronous network of classical machines that send \( b \) classical bits per time unit? 
It turns out that in many cases, the answer is \emph{yes}. 
There are computational tasks that are asymptotically easier to solve in the \qcongest model (and related variants): 
see, e.g., \cite{izumi2019,izumi2020,legall2018}.

However, if we consider instead networks that are constrained only by \emph{large distances}, the scenario is much less understood. 
Such networks typically model distributed systems where network latency and the time required for information to propagate play key roles. We emphasize that large distances are a fundamental physical limitation of distributed networks: 
information, whether classical or quantum, cannot travel faster than the speed of light. 
This limitation cannot be overcome by technological progress, unlike bandwidth constraints, which can benefit from innovations (say, the installation of multiple parallel communication channels by, e.g., increasing the number of fiber-optic links between nodes).

Such distance-constrained networks are modeled by the famous \local model of computation, first introduced in the seminal work by \textcite{linial87} (see \cite{linial92} for the journal version). 
In the \local model, a distributed network is represented as a graph \(G = (V, E)\), where the nodes of \(G\) are processors capable of \emph{unbounded local computation}, and the edges represent communication links. 
Time proceeds synchronously: 
in each round, nodes send and receive messages of \emph{arbitrarily large size} from their neighbors and perform local computations to update their state variables. 
Eventually, all nodes announce their outputs, marking the end of the computation. 
The complexity measure in this model is the number of communication rounds required to solve a problem, captured by the notion of \emph{locality}. 
Specifically, \(T\) rounds of communication allow a node to gather information about the topology and input within its radius-\(T\) neighborhood. 
Thus, \(T\) is also called the \emph{locality} of the algorithm, reflecting why distances are the only limitation in this model.

The \local model has been extensively studied \cite{linial92,naor1991,naor1995,chang16exponential,chang_kopelowitz_pettie2019exp_separation,gavoille2009,chang19hierarchy,ghaffari2018derandomizing,fischer_ghaffari2017sublogarithmic,ghaffari2017,brandt16lll,hirvonen14local-maxcut,kuhn2004,korman-2011-global-knowledge}.
A specific class of problems of particular interest in distributed computing is \emph{locally checkable labeling} (LCL) problems, introduced by \textcite{naor1995}. 
LCL problems are defined via local constraints (e.g., graph coloring). 
For LCL problems, a solution may be \emph{hard to find} but is \emph{easy to verify} with a distributed algorithm.
As such, LCL problems can bee seen as the distributed analogue of the FNP class in centralized computation. 
Nowadays, we have a good understanding of complexity landscape of LCL problems in the \local model \cite{naor1995,chang_kopelowitz_pettie2019exp_separation,chang16exponential,chang19hierarchy,balliu18lcl-complexity,balliu19lcl-decidability,balliu19mm,balliu20almost-global,balliu20lcl-randomness,balliu21lcl-congest,balliu22mending,balliu22regular-trees,balliu22rooted-trees,balliu22so-simple,brandt17grid-lcl,Brandt2019automatic,akbari_et_al:LIPIcs.ICALP.2023.10}.

To date there is no clear understanding of the impact of quantum computation and communication in the \local model, especially concerning LCL problems. 
The main challenge lies in the lack of tools for directly tackling \qlocal, particularly for establishing lower bounds. 
Nevertheless, general arguments based on physical principles, such as \emph{causality} and \emph{independence} of output distributions, provide some insights. 

In this brief survey, we summarize previous knowledge and recent results, introducing new techniques for investigating the role of quantum computation and communication in distributed settings and shedding light on potential research directions.
\section{Preliminaries}\label{sec:preliminaries}

In order to proceed, we need to provide the mathematical framework in which we work. 
We start with some basic graph notations and definitions, introducing the class of problems we consider and the computational model that is the heart of our investigation.
Section by section, we will give other definitions that are useful for that section and the subsequent ones.
We denote the set of natural numbers (starting from 0) by \(\nats\), and also define \(\natsPos = \nats \setminus \{0\}\).

\paragraph{Graphs.} We work with simple graphs unless otherwise specified.
Given any set \(S\), we denote by \(\binom{S}{k}\) the set whose elements are all sets of \(k\) different elements of \(S\).
We will consider both undirected and directed graphs.
A graph is a pair \(G = (V,E)\) where \(V\) is the set of nodes and \(E\) is the set of edges.
In case of undirected graphs, \(E \subseteq \binom{V}{2}\), whereas \(E \subseteq V \times V\) in case of directed graphs.
For any graph \(G\), we also denote its set of nodes by \(V(G)\) and its set of edges by \(E(G)\).

The distance between two nodes \(u,v\) of any graph \(G\) is the number of edges in any shortest path between \(u\) and \(v\) (note that the shortest path is not necessarily an oriented path), and is denoted by \(\dist_G(u,v)\).
The notion of distance can be easily extended to subset of nodes: 
Given any node \(u \in V\) and any two subsets \(A,B \subseteq V\), the distance between \(u\) and \(A\) is \(\dist_G(u,A) = \min_{v \in A}\{\dist_G(u,v)\}\) and the distance between \(A\) and \(B\) is \(\dist_G(A,B) = \min_{u \in A, v \in B}\{\dist_G(u,v)\}\).
When the graph is clear from the context, we omit the suffix and write only \(\dist()\) instead of \(\dist_G()\).
Through the notion of distance, we define the diameter of a graph \(G\) to be \(\diam(G) = \max_{u,v \in V(G)} \dist_G(u,v)\).

For any non-negative integer \(T\), the radius-\(T\) (closed) neighborhood of a node \(u\) in a graph \(G\) is the set \(\neighborhood_T[u] = \left\{v \in V \st \dist(u,v) \le T \right\}\).
Throughout this survey, we will only make use of closed neighborhoods.
More in general, the radius-\(T\) neighborhood of any subset of node \(A \subseteq V\) is \(\neighborhood_T[A] = \cup_{u \in A} \neighborhood_T[u]\).
We also define the \emph{ring neighborhood} between \(T_1\) and \(T_2\) of a node \(u \in V(G)\) (for \(T_1 \le T_2\)) as \(\neighborhood_{T_2}^{T_1}[u] = \neighborhood_{T_2}[u] \setminus \neighborhood_{T_1}[u]\). 
We use an analogous notation for subsets of nodes.

The degree of a node \(v\) in an undirected graph \(G\) is the number of edges the node belongs to, i.e., \(\deg_G(v) = \abs{\left\{ \{u,v\} \in E(G) \st u \in V(G)\right\}}\).
In a directed graph \(G\), we define the \emph{indegree} and the \emph{outdegree} of a node \(v\) as follows:
\(\indeg_G(v) = \abs{\left\{(u,v) \in E(G) \st \ u \in V(G)\right\}}\) and \(\outdeg_G(v) = \abs{\left\{(v,u) \in E(G) \st \ u \in V(G)\right\}}\).
Then the degree of \(v\) in \(G\) is just \(\deg_G(v) = \indeg_G(v) + \outdeg_G(v)\).
In all these notations, we omit the suffix when the graph is clear from the context.
Finally, the degree of a graph \(G\) is just the maximum degree of any node, i.e., \(\deg(G) = \max_{v \in V(G)} \{\deg_G(v) \}\).

For any graph \(G = (V,E)\) and any subset of nodes \(A \subseteq V\), the subgraph of \(G\) induced by \(A\) is denoted by \(G[A]\).
Consider any node \(u \in V\) (or any subset \(S \subseteq V\)): with an abuse of notation, we define the \emph{open induced subgraph} as the set \(\mathring{G}[\neighborhood_{T}[u]] = G[\neighborhood_{T}[u]] \setminus G[\neighborhood_{T}^{T-1}[u]]\). (or \(\mathring{G}[\neighborhood_{T}[S]] = G[\neighborhood_{T}[S]] \setminus G[\neighborhood_{T}^{T-1}[S]]\)):.
In practice, in this definition we are removing from the classical notion of neighborhood the edges that connect nodes that are at distance \(T\) from \(u\) (or \(S\)) as the graph that \(u\) (or \(S\)) \emph{sees} by moving \(T\) hops away does not include them.

Now we define some graph operations. 
Given any two graphs \(G\) and \(H\), the intersection of \(G\) and \(H\) is the graph \(G \cap H = (V(G) \cap V(H), E(G) \cap E(H))\).
The union of \(G\) and \(H\) is the graph \(G \cup H = (V(G) \cup V(H), E(G) \cup E(H))\), while the difference between \(G\) and \(H\) is the graph \(G \setminus H = (V(G) \setminus V(H), E(G) \setminus E(H)) \).

An isomorphism between two graphs \(G\) and \(H\) is a function \(\varphi: V(G) \to V(H)\) that is bijective and such that \(\{u,v\} \in E(G)\) if and only if \(\{\varphi(u), \varphi(v)\} \in E(H)\).
In case of directed graphs, we also require the isomorphism to keep edge orientation.

\paragraph{Labeling problems.} 
In this brief survey we consider labeling problems, namely, graph problems that ask to output some labels on the nodes of the graph.
A formal definition follows.
\begin{definition}[Labeling problem]\label{def:preliminaries:labeling-problem}
    Let \(\inputSet,\outputSet\) be two alphabets, and \(I\) a set of indices.
    A labeling problem \(\problem\) is a mapping \((G,\inpt) \mapsto \{\oupt_i\}_{i \in I}\) that maps every input graph \(G = (V,E)\) where nodes are labelled by any input function \(\inpt: V \to \inputSet\) to a family of suitable output functions \(\oupt_i : V \to \outputSet\) indexed by \(I\).
    The mapping is closed under graph isomorphism, that is, for any isomorphism \(\varphi: V(G) \to V(H)\) between two graphs \(G\) and \(H\), \(\oupt \in \problem((H,\inpt)) \) if and only if \(\oupt \circ \varphi \in \problem((G,\inpt \circ \varphi))\). 
\end{definition}
The reader may notice that we defined labeling problems for \emph{any input graph}, despite the fact that some labeling problems might be defined only for some specific graph family like, for example, \(3\)-coloring bipartite graphs.
However, \cref{def:preliminaries:labeling-problem} is general enough to capture all such problems, because one can just say that, for graphs that are outside the right graph family (e.g., non-bipartite graphs), all outputs are admissible.
Examples of graph labeling problems are leader election, consensus, diameter approximation, etc.

A subclass of labeling problems that is of particular interest in the distributed computing community is that of locally checkable labeling (LCL) problems, already mentioned in the introduction.
LCL problems were first introduced by \textcite{naor1993} (for the journal version we refer to \cite{naor1995}).
Here, we report the original definition.

For any function \(f: A \to B\) and any subset \(S \subseteq A\), we denote the restriction of \(f\) to \(S\) by \(f \restriction_S : S \to B\).
We define a \emph{centered graph} to be any pair \((G, v_G)\), where \(G\) is a graph and \(v_G \in V(G)\) is a node of \(G\) that we call the \emph{center} of \(G\).
The \emph{radius} of a centered graph \((G,v_G)\) is the maximum distance between \(v_G\) and any other node of \(G\).
We are now ready to state the definition of LCL problems.

\begin{definition}[Locally Checkable Labeling problem]\label{def:preliminaries:lcls}
    Let \(r\), \(\maxDeg\) be non-negative integers.
    Let \(\inputSet, \outputSet\) be two finite alphabets, and \(I\) a finite set of indices. 
    Consider a labeling problem \(\problem\) defined on \(\inputSet, \outputSet, I\).
    \(\problem\) is \emph{locally checkable} with checking radius \(r\) and maximum degree \(\maxDeg\) if there exists a family \( \SS = \left\{ \left((H, v_H), \inpt, \oupt\right)_i\right\}_{i \in I}\) where each tuple \(\left((H, v_H), \inpt, \oupt\right)_i\) contains a centered graph \((H,v_H)\) of radius at most \(r\) and degree at most \(\maxDeg\), an input labeling function \(\inpt: V(H) \to \inputSet\) and an output labeling function \(\oupt: V(H) \to \outputSet\) with the following property:
    \begin{itemize}
        \item For every input \((G, \inpt)\) to \(\problem\) with \(\deg(G) \le \maxDeg\), an output vector \(\oupt:V(G) \to \outputSet\) is admissible (i.e., \(\oupt \in \problem((G,\inpt))\)) if and only if, for each node \(v \in V(G)\), the tuple \(\left((G[\neighborhood_{T}[v]], v), \inpt\restriction_{\neighborhood_{T}[v]}, \oupt\restriction_{\neighborhood_{T}[v]}\right)\) belongs to \(\SS\).
    \end{itemize}
\end{definition}

\(\SS\) is also called the family of \emph{permissible outputs}, and is finite (up to graph isomorphisms) since \(\maxDeg,r\) are finite and also \(\inputSet,\outputSet\) are finite sets.
Notice that in \cref{def:preliminaries:labeling-problem,def:preliminaries:lcls} we assumed that input and output labels are given to and from nodes. 
We might similarly assume that they are given to and from edges, and come up with other problems. 
In this survey, we will interchangeably use both possibilities when it is convenient.
LCL problems capture all problems defined via local constraints, e.g., graph coloring, maximal independent set, maximal matching, sinkless orientation, triangle freeness, etc. 

We now introduce the models of computations we are interested in.
We begin with the \portnum model, and on top of it we define the \local model of computation \cite{linial92}.

\paragraph{The \portnum model.}
A port-numbered network is a triple \(N = (V, P, p)\) where \(V\) is the set of nodes, \(P\) is the set of \emph{ports}, and \(p: P \to P\) is a function specifying connections between ports.
Each element \(x\in P\) is a pair \((v,i)\) where \(v \in V\), \(i \in \natsPos\).
The connection function \(p\) between ports is an involution, that is, \(p(p(x)) = x\) for all \(x \in P\).
If \((v,i) \in P\), we say that \((v,i)\) is port number \(i\) in node \(v\).
With an abuse of notation, we say that the degree of a node \(v\) in the network \(N\) is the number of ports in \(v\) and is denoted by \(\deg_N(v)\).
We assume that port numbers are consecutive, i.e., the ports of any node \(v \in V \) are \((v,1), \dots, (v, \deg_N(v))\).
Clearly, a port-numbered network identifies an \emph{underlying graph} \(G = (V,E)\) where, for any two nodes \(u,v \in V\), \(\{u,v\} \in E\) if and only if there exists ports \(x_u,x_v \in P\) such that \(p(x_u) = x_v\).
Here, the degree of a node \(\deg_N(v)\) corresponds to \(\deg_G(v)\).

In the \portnum model we are given a distributed system consisting of a port-numbered network of \( \abs{V} =
n\) \emph{processors} (or \emph{nodes}) that operates in a sequence of
synchronous rounds.
In each round the processors may perform unbounded computations on their
respective local state variables and subsequently exchange messages of
arbitrary size along the links given by the underlying input graph.
Nodes identify their neighbors by using ports as defined before, where the port assignment may be done adversarially.
Barring their degree, all nodes are identical and operate according to the
same local computation procedures.
Initially all local state variables have the same value for all processors;
the sole exception is a distinguished local variable \(\localVar(v)\) of
each processor \(v\) that encodes input data (that is, port numbers, degree, possible input from the problem itself, etc.).
Usually, we assume that \(\localVar(v)\) also encodes the number of nodes \(n\) composing the distributed system.

Let \(\inputSet\) be a set of input labels.
The input of a problem is defined in the form of a labeled graph \((G,
\inpt)\) where \(G = (V, E)\) is the system graph, \(V\) is the set of
processors (hence it is specified as part of the input), and \(\inpt\colon V
\to \inputSet\) is an assignment of an input label \(\inpt(v) \in \inputSet\) to
each processor \(v\) and is encoded in \(\localVar(v)\).
The output of the algorithm is given in the form of a function of output labels
\(\oupt\colon V \to \outputSet\), and the algorithm is assumed to
terminate once all labels \(\oupt(v)\) are definitely fixed.
We assume that nodes and their links are fault-free.
The local computation procedures may be randomized by giving each processor
access to its own set of random variables; in this case, we are in the
\emph{randomized} \portnum model as opposed to the \emph{deterministic}
\portnum model.

The running time of an algorithm is the number of synchronous rounds required by all nodes to produce output labels.
If an algorithm running time is \(T\), we also say that the algorithm has locality \(T\).
Notice that \(T\) can be a function of the size (or other parameters) of the input graph. 
We say that a problem \(\problem\) over some graph family \(\FF\) has complexity (or locality) \(T\) in the \portnum model if there is a \portnum algorithm running in time \(T\) that solves \(\problem\) over \(\FF\), and \(T=T(n)\) is the minimum running time (among all possible algorithms that solve \(\problem\) over \(\FF\)) in the worst case instance of size \(n\).
If the algorithm is randomized, we also require that the failure probability is at most \(1/\poly(n)\), where \(n\) is the size of the input graph.

We remark that the notion of an (LCL) problem is a graph problem, and does not depend on the specific model of computation we consider (hence, the problem definition cannot depend on, e.g., port numbers).

\paragraph{The \local model.}

The \local model of computation is just the \portnum model augmented with an assignment of unique identifiers to nodes.
Let \(c \ge 1\) be a constant.
The nodes of the input graph \(G = (V,E)\) are given as input also unique identifiers specified by an injective function \(\text{id}: V \to [n^c] \).
This assignment might be adversarial and is stored in the local state variable \(\localVar(v)\), and nodes can exploit these values during their local computation.

The local computation procedures may be randomized by giving each processor
access to its own set of random variables; in this case, we are in the
\emph{randomized} \local (\randlocal) model as opposed to the \emph{deterministic}
\local (\detlocal) model.
If the algorithm is randomized, we also require that the failure probability while solving any problem is at most \(1/\poly(n)\), where \(n\) is the size of the input graph.
The definition of running time, locality and complexity easily extends from the \portnum model to the \local model.

On top of the \local model, it is easy to describe its quantum generalization. 
In order to avoid the math of quantum mechanics, we only provide an informal definition of the \qlocal model.
For a formal definition, we defer the reader to \cite{gavoille2009}.

\paragraph{The quantum-LOCAL model.}
The \qlocal of computing is similar to the deterministic \local model above, but now with quantum computers and quantum communication links. More precisely, the quantum computers manipulate local states consisting of an unbounded number of qubits with arbitrary unitary transformations, the communication links are quantum communication channels (adjacent nodes can exchange any number of qubits), and the local outputs can be the result of any quantum measurement.

\begin{figure}[h!]
    \centering
    \includegraphics[scale=1]{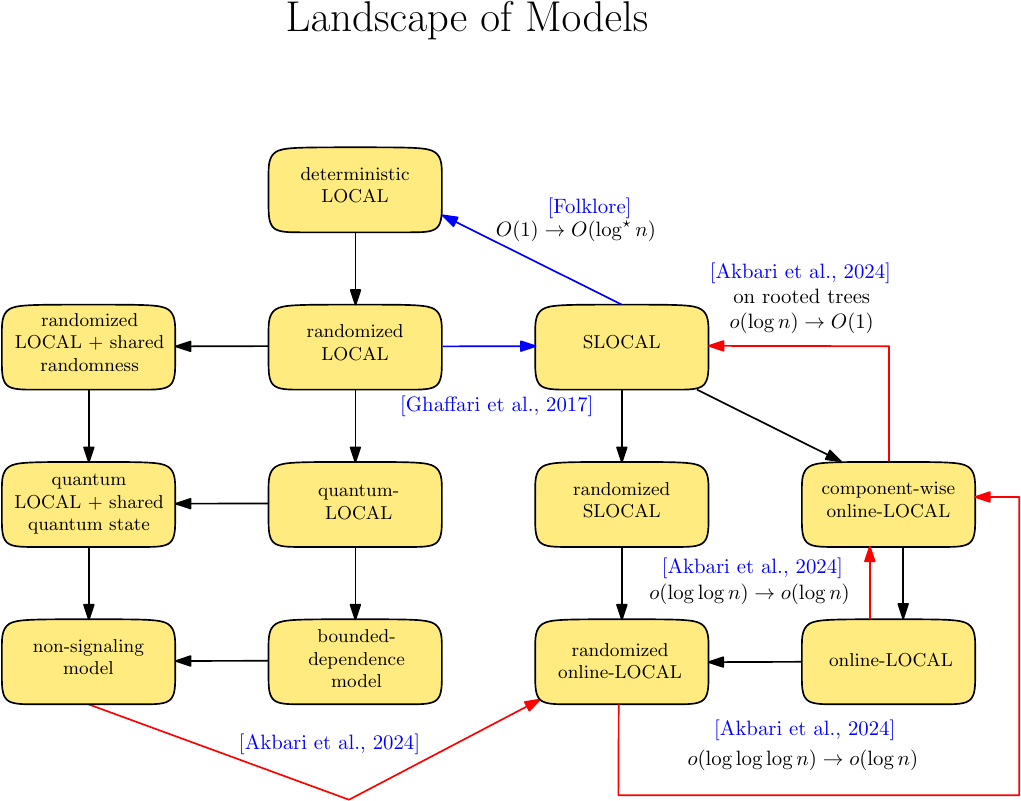}
    \caption{Landscape of computational models. 
    An arrow between model \(X\) and \(Y\), that is \(X \to Y\) means that model \(Y\) is stronger than model \(X\), unless otherwise specified.
    Black arrows are trivial implications (by construction), blue arrows are known results, and red arrows are recent results.}
    \label{fig:landscape-of-models}
\end{figure}

\paragraph{Relations between models.}
We say that a computational model \(A\) is \emph{stronger} than a computational model \(B\) if an algorithm with locality \(T\) running in \(A\) can be simulated by an \(\myO{T}\)-round algorithm in \(B\).
Clearly, \detlocal is stronger than the \portnum model, \randlocal is stronger than \detlocal, and \qlocal is stronger than \randlocal.
We suggest the reader has \cref{fig:landscape-of-models} at hand to keep track of the models and their relations while we introduce them.
We will define the other models present in \cref{fig:landscape-of-models} later in the related sections.

\section{The \nonsign model}
\label{sec:non-signaling}

As mentioned in the introduction, to date we do not have direct ways to prove lower bounds in \qlocal.
Specific procedures that are commonly used in the \local model such as round elimination \cite{Brandt2019automatic} do not generalize to \qlocal (regarding this, we argue more later in \cref{sec:quantum-advantage:iteratedGHZ}).
However, some general arguments based on the \emph{\nonsign} principle do generalize to \qlocal and actually holds more in general.
Before going through specific definition, let us introduce the no-signaling principle via two examples on the same problem: 2-coloring even cycles both in \detlocal and in \randlocal.

\subsection{Warm-up: 2-coloring cycles in classical \local}\label{sec:non-signaling:warm-up}

Let \(n \in \nats\) be an even number and consider a cycle \(C_n\) with \(n\) nodes.

\begin{figure}[p]
    \centering
    \begin{subfigure}{\textwidth}
      \centering
      \includegraphics[scale=0.66625]{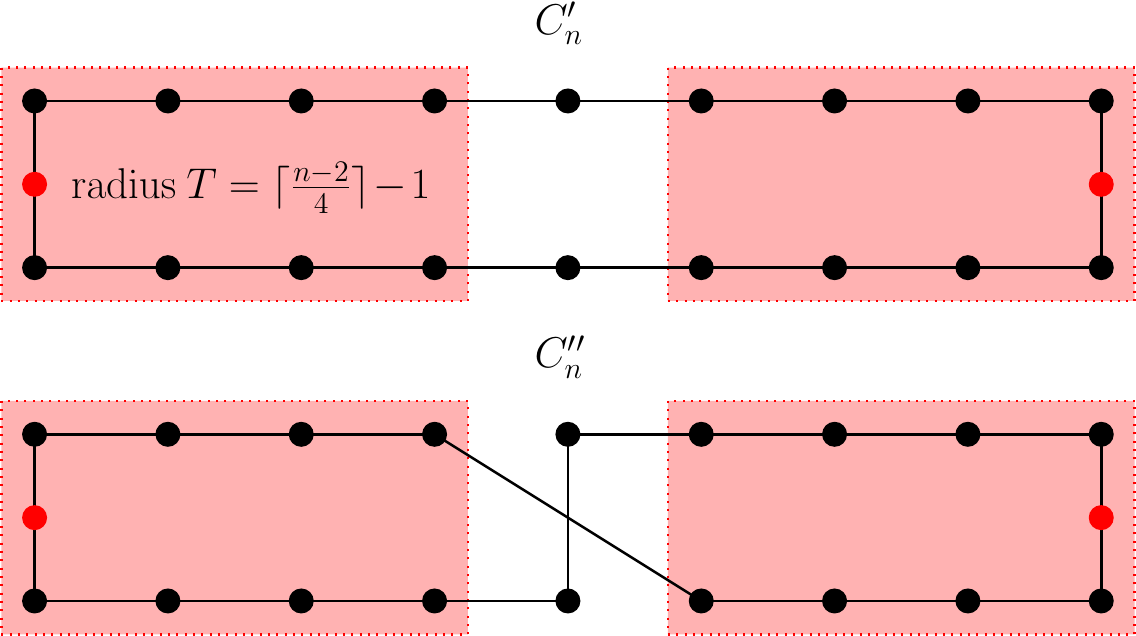}
      \caption{Indistinguishability argument \emph{within} the same input graph family.
      The two highlighted nodes cannot distinguish between the case in which they have even (in \(C'_n\)) or odd (in \(C''_n\)) distance between each other.}
      \label{fig:non-signaling:indistinguishability-cycles}
    \end{subfigure}
    \\[5mm]
    \begin{subfigure}{\textwidth}
      \centering
      \includegraphics[scale=0.66625]{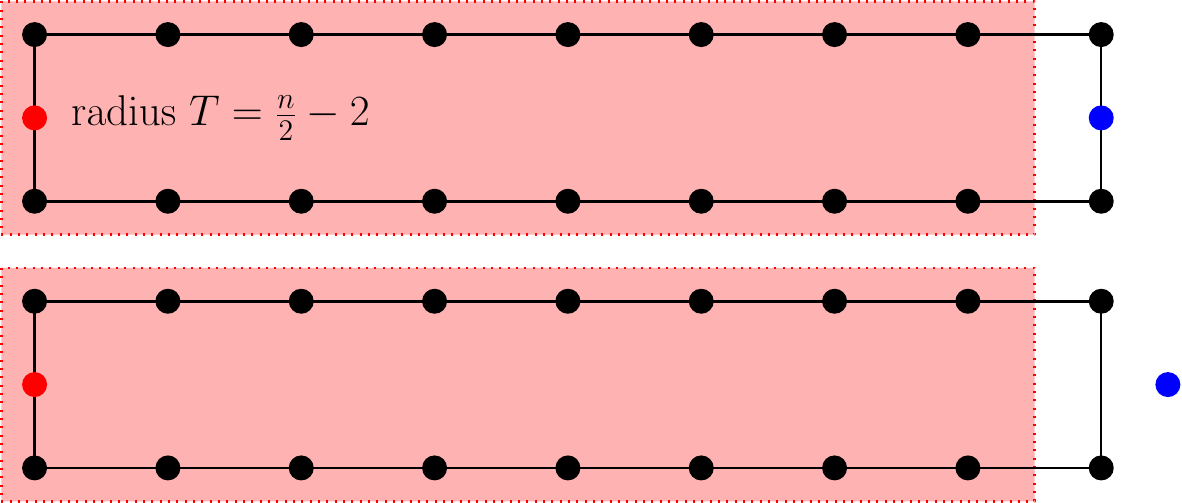}
      \caption{Graph-existential argument: \emph{outside} the input graph family.
      The red nodes cannot distinguish whether they are in an even or an odd cycle.}
      \label{fig:non-signaling:graph-existential-cycles}
    \end{subfigure}
    \caption{The no-signaling principle illustrated in the problem of 2-coloring cycles.}
    \label{fig:non-signaling:cycles}
  \end{figure}

\paragraph{First example: indistinguishability argument (\emph{within} the input graph family).}
The indistinguishability argument in the \local model relies on the fact that a node, by accessing its local view, cannot distinguish between \emph{proper} inputs that are indistinguishable in the local neighborhood but differ outside and, for which, the node must behave in different ways. 
For example, the bipartitness of an even cycle is a very rigid property: if the radius-\(T\) neighborhoods of two distinct nodes in a cycle do not intersect, the nodes cannot guess if they are in the same set of the bipartition or not.
More formally, assume that there is a \detlocal algorithm \(\algo\) that \(2\)-colors \(C_n\) in time \(T = \ceil{(n-2)/4} -1\).
Consider two nodes \(u_1,u_2\) that are at distance at least \(2\ceil{(n-2)/4} - 1\) between each other.
Their radius-\(T\) neighborhoods do not intersect, hence there is no way to coordinate in case their distance is odd (which implies different colors) or even (which implies the same color).
One can instantiate two input graphs \(C_n'\) and \(C_n''\) in which the distance between \(u_1\) and \(u_2\) is even and odd, respectively. 
However, when running \(\algo\) the output must be the same, leading to a contradiction (see \cref{fig:non-signaling:indistinguishability-cycles}).
This argument gives a lower bound of \(\ceil{(n-2)/4} -1\) to the problem.
As for \randlocal, simply notice that, in the worst case, the two nodes \(u_1\) and \(u_2\) cannot produce the correct output with probability more than \(1/2\).

\paragraph{Second example: graph-existential argument (\emph{outside} the input graph family).}
Graph-existential arguments are another key technique to prove lower bounds in classical \local.
The idea is the following: Suppose we have an LCL problem that assume that the input graph family satisfies some key-property (in this case, bipartite cycles).
Assume also that one can find a graph that \emph{locally} looks like a proper input, but lies \emph{outside} the input graph family (e.g., an odd cycle).
In this way, one can exploit the fact that, if the locality \(T\) of the algorithm solving the problem is not large enough to detect that the graph is not proper, then the algorithm must run and produce some local failure (e.g., two monochromatic nodes).
We can now take a copy of the radius-\(T\) neighborhood of the failing nodes and construct a proper input that contains this copy. 
Since the nodes do not distinguish in which input (proper or not) they are in, they must produce the same output, leading to the failure of the algorithm in a proper instance (see \cref{fig:non-signaling:graph-existential-cycles}).
Formally, assume that there is a \detlocal algorithm \(\algo\) that \(2\)-colors \(C_n\) in time \(T = n/2 - 2\).
Consider now a second graph \(G\) of \(n\) nodes that is the disjoint union of a cycle \(C_{n-1}\) with \(n-1\) nodes and a single node with no edges.
Run \(\algo\) in \(G\): the nodes in \(C_{n-1}\) will not notice that they are in an odd cycle since the radius-\(T\) neighborhood of any node leaves out \(2\) nodes.
Hence, the nodes must output some color and, since \(C_{n-1}\) is not 2-colorable, there must be a failure somewhere, that is, two adjacent nodes \(u_1,u_2\) share the same color.
We can now construct an even cycle \(C_n\) that contains a subgraph isomorphic to \(C_{n-1}[\neighborhood_{T}[\{u_1,u_2\}]]\), with the same input data (that is, identifiers and port numbers).
Since the isomorphic copies of \(u_1,u_2\) in \(C_n\) cannot distinguish if they are in \(C_n\) or in \(C_{n-1}\) when running \(\AA\), they must produce the same failure, contradicting the hypothesis that \(\AA\) was correct.
This argument gives a lower bound of \(n/2 - 2\) for \(2\)-coloring even cycles with \(n\) nodes.

When we allow randomness, things are a bit more complex.
Indeed, the failure in \(C_{n-1}\) is random and when we look at two specific adjacent nodes, in the worst case, the probability of a failure (i.e., a monochromatic edge) can be as small as \(1/\poly(n)\).
However, we can do something different.
Let \(u_1, \dots, u_{n-1}\) be the nodes of \(C_{n-1}\), where \(\{u_{i}, u_{i+1}\}\) is an edge for \(i = 1, \dots, n-2\), and \(\{u_1, u_{n-1}\}\) is another edge.
Consider two subgraphs \(G,H\) of \(C_{n-1}\) defined as follows: \(V(G) = \{u_1, \dots, u_{n/2}\}\), \(V(H) = \{u_{n/2}, \dots, u_{n-1}\}\).
Now assume \(\AA\) is a \randlocal algorithm that \(2\)-colors even cycles of \(n\) nodes in time \(T = \ceil{(n-2)/4} - 1\), and run \(\AA\) in \(C_{n-1}\) (plus one disjoint singleton node to make the number of nodes equal to \(n\)).
Again, since locality \(T\) is not sufficient for the nodes of \(C_{n-1}\) to understand that they are not in an even cycle, there must be a failure in at least two adjacent nodes \(u,v\) at every run of \(\AA\).
Since \(G \cup H = C_{n-1}\), in at least one of them there is probability no less than \(1/2\) that a failure takes place.
Wlog, assume \(G\) is such subgraph.
Then, one can construct an even cycle of \(n\) nodes that contains as induced subgraph a copy of \(\mathring{C}_{n-1}[\neighborhood_{T}[V(G)]]\) and give as input the same identifiers and port numbers.
Since the view of the nodes of the copy of \(G\) in \(C_n\) is indistinguishable  from the view of the nodes of \(G\) in \(C_{n-1}\), they must reproduce the same failure probability.
Hence, we get that any algorithm that \(2\)-colors cycles in \randlocal with locality at most \(T = \ceil{(n-2)/4} - 1\) fails with probability at least \(1/2\).
Notice that we could consider more subgraphs of \(C_n\) and get lower bounds with higher locality but smaller failure probability.

\begin{figure}[h!]
    \centering
    \includegraphics[scale=0.7]{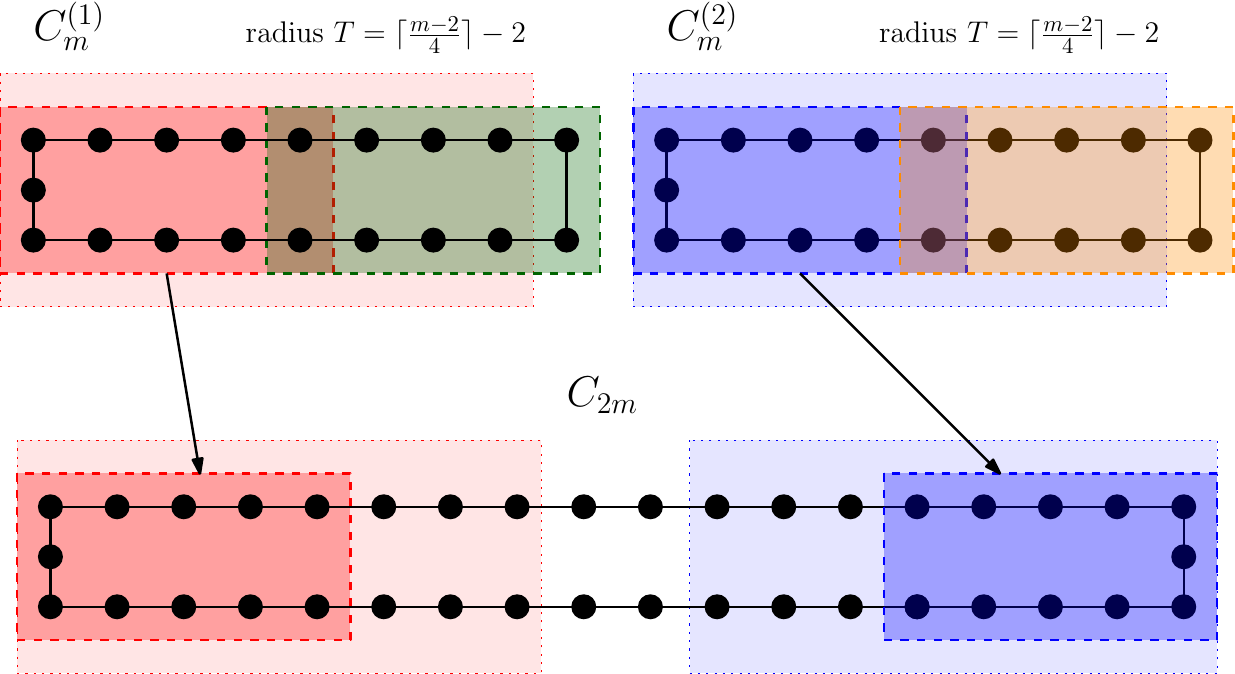}
    \caption{Visual representation of the failure probability boost: We take two odd cycles \(C^{(1)}_m\) and \(C^{(2)}_m\) and subdivide the nodes in two slightly overlapping regions (the colored dashed regions).
    For \(T = \ceil{\frac{m-2}{4}}-2\), a \(T\)-round algorithm \(\algo\) must fail both in \(C^{(1)}_m\) and \(C^{(2)}_m\) (it does not catch that we are in odd cycles).
    For each cycle, in at least one region \(\algo\) must fail with probability at least 1/2.
    Wlog, we assume that this happens in the red and the blue regions.
    Now we create a new cycle \(C_n\) with \(n = 2m\) nodes where we copy the radius\(T\) neighborhoods of the red and the blue region, and we add the remaining nodes.
    The failing probability of \(\algo\) over \(C_n\) is, by independence, at least \(3/4\), and we get a lower bound on the locality of magnitude \(T = \ceil{\frac{n-4}{8}} - 2\).}    
    \label{fig:non-signaling:cycles-boosting}
\end{figure}

\paragraph{Randomized \local: boosting the failure probability.} 
In \randlocal, we can also boost the failure probability at the cost of worsening the locality of the lower bound, by simply repeating the experiment many times.
We only focus on the graph-existential argument, but a similar approach holds for the other argument as well.
We assume that the overall amount of nodes is now \(n = m \cdot N\) for two positive integers \(m, N\) where \(m\) is odd and \(N\) is even.
Assume now that we have an algorithm \(\AA\) that \(2\)-colors even cycles of \(n\) nodes with locality \(T = \ceil{(m-2)/4}-1 = \ceil{(n/N - 2)/4}-1\).
Now consider \(N\) disjoint copies \(C_m^{(1)},\dots,C_m^{(N)} \)of an odd cycle \(C_m\) with \(m\) nodes.
For each \(C_m^{(i)}\), the same argument as before holds, and we can identify a subgraph \(G_i\) of each \(C_m^{(i)}\) where a failure takes place with probability at least \(1/2\) independently of the others, and such that the radius-\(T\) neighborhood of \(V(G)\) still leaves at least one node of \(C_m^{(i)}\) out.
Now we can construct a proper input \(C_n\) that contains, as induced subgraphs, a
copy of each \(C_m^{(i)}[\neighborhood_{T}(V(G_i))]\).
By independence, the failure probability here is at least \(1 - \frac{1}{2^N}\).
Hence, any algorithm \(2\)-coloring even cycles with locality \(T = \ceil{(n/N - 2)/4}-1\) fails with probability at least \(1 - \frac{1}{2^N}\).
See \cref{fig:non-signaling:cycles-boosting} for a visual explanation.

\subsection{The no-signaling principle}

The lower bound techniques in \cref{sec:non-signaling:warm-up} rely on a crucial assumption, which is quite intuitive when dealing with classical \local.
First, let us introduce the notion of \emph{view} of a subset of nodes.

For any input distributed network \((G,\localVar)\) to any problem, and any subset of nodes \(S\subseteq V(G)\), the radius-\(T\) view of \(S\) is \(\view_T(S) = (\mathring{G}[\neighborhood_{T}[S]], \localVar \restriction_{\neighborhood_{T}[S]})\).
Basically, \(\view_T(S)\) includes everything that can be \emph{seen} by the nodes in \(S\) with \(T\) rounds of communication, including input data (degree, ports, identifiers and input labels---if any, etc.).
Suppose \(G\) has \(n\) nodes, and fix any subset of nodes \(S \subseteq V(G)\). 
Given any two graphs \(G,H\) with inputs \(\localVar_G, \localVar_H\) and any two subset of nodes \(S_G \subseteq V(G)\) and \(S_H\subseteq V(H)\), it is natural to define the notion of \emph{isomorphism between views}.
We say that
\(\view_T(S_G)\) is isomorphic to \(\view_T(S_H)\) if there exists a function \(\varphi: V(G) \to V(H)\) such that the folowwing holds: 
\begin{enumerate}
    \item \(\varphi \restriction _{S_G}\) is an isomorphism between \(G[S_G]\) and \(H[S_H]\)
    \item \(\varphi \restriction _{\NN_T[S_G]}\) is an isomorphism between \(\mathring{G}[\NN_T[S_G]]\) and \(\mathring{H}[\NN_T[S_H]]\);
    \item \(\localVar_G(u) = \localVar_H(\varphi(u))\).
\end{enumerate}

\begin{figure}[h!]
    \centering
    \includegraphics[scale=0.8]{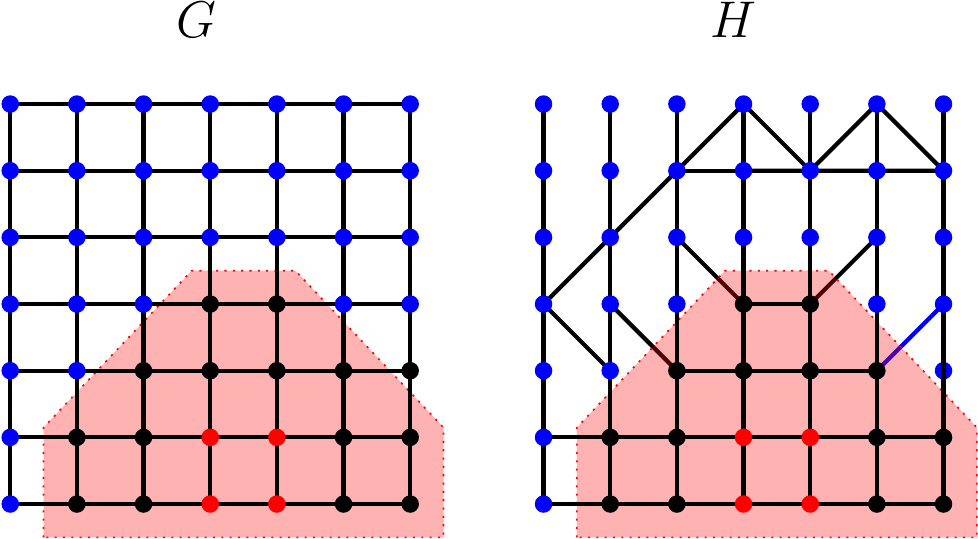}
    \caption{No-signaling property.
    Alice's lab contains the red nodes, Bob's lab contains the blue nodes.
    By running any \(2\)-rounds synchronous distributed algorithm, the red nodes cannot distinguish between \(G\) and \(H\): Bob has freedom to change the topology outside the red region without being detected by Alice.}
    \label{fig:non-signaling:property}    
\end{figure}

Consider any \(T\)-round (deterministic or randomized) \local algorithm \(\algo\) run by the nodes of \(G\).
Imagine that the distributed network is split in two different laboratories, Alice's and Bob's. 
Alice's lab contains \(\view_0(S)\), while Bob's lab contains everything that is not contained in \(\view_T(S)\).
When performing \(T\) rounds of communication, Alice's lab receive no information from Bob's lab. 
Hence, the behavior of the output distribution over nodes in Alice's lab cannot change whatever Bob does in his lab, including rearranging links between nodes or manipulating inputs.
The \emph{cause} of outputs in Alice's lab \emph{cannot be influenced} by any action of Bob in his lab.
See \cref{fig:non-signaling:property} for a visual representation of the property.\footnote{Formally, Bob's lab also contains edges that are in \(E(G[\neighborhood_T[S]]) \setminus E(\mathring{G}[\neighborhood_T[S]])\), but does not contain the nodes in \(\neighborhood_{T}^{T-1}[S]\).
We can imagine that Bob sees the ports of edges that are not in \(\view_T(S)\), but nothing else, and is constrained to assign edges to those ports.
In \cref{fig:non-signaling:property} Bob would have the freedom to modify such edges in the red region as well, but we avoid representing this aspect for the sake of simplicity.}
This property is the so-known \emph{no-signaling from the future} principle in physics, which states that no signals can be sent from the future to the past, and is equivalent to the \emph{causality principle} \cite{d2017quantum}. 
Such principle holds in \emph{every physical distributed network} running any kind of synchronous distributed algorithm, including quantum ones.

To see how this principle formally translates in our setting, let us define the notion of \emph{outcome}, which is some kind of generalization of an algorithm.
Here, we restrict to finite sets of input and output labels since we focus on LCL problems.
\begin{definition}[Outcome]
    Let \(\inputSet,\outputSet\) be two finite sets of labels, and \(I\) a finite set of indices.
    An \emph{outcome} is a mapping \(\outcome: (G, \localVar) \mapsto \{(\oupt_i: V(G) \to \outputSet, p_i)\}_{i \in I}\) that assigns to every input distributed network \((G, \localVar)\) (with any input labeling \(\inpt: V(G) \to \inputSet\)) a distribution over output labelings \(\{(\oupt_i: V(G) \to \outputSet, p_i)\}_{i \in I}\), where \(p_i\) is the probability that \(\oupt_i\) occurs; in particular, \(0 \le p_i \le 1\) and \(\sum_{i \in I} p_i = 1\).
\end{definition}
This definition is easily generalizable to the case of infinite label sets, but we avoid it for the sake of simplicity.
Notice that all synchronous distributed algorithms gives r outcomes: it is just the assignment of an output distribution (the one that is the result of the algorithmic procedure) to the input graph.
We remark we have defined the domain set of outcomes to include \emph{all possible graphs}. 
This is not restrictive: Even classical (or quantum) algorithms can run on every possible graph. 
One can just introduce some garbage output label so that whenever a node running an algorithm needs to do something that is not well-defined (given its neighborhood), it can just output the garbage label.

We say that an outcome \(\outcome\) solves a problem \(\problem\) over a family of graphs \(\FF\) with probability \(q > 0\) if, for every \(G \in \FF\) and every input data \(\localVar\), it holds that
\[
    \sum_{\substack{i \in I : \\ \oupt_i \in \problem((G,\localVar))}} p_i \ge q.
\]
Let \(\outcome: (G, \localVar) \mapsto \{(\oupt_i, p_i)\}_{i \in I}\) be any outcome and fix an input \((G, \localVar)\).
Consider any subset of nodes \(S \subseteq V(G)\).
The restriction of the output distribution \(\outcome((G,\localVar)) = \{(\oupt_i: V(G) \to \outputSet, p_i)\}_{i \in I}\) to \(S\) is the distribution 
\(\{(\oupt_j: S \to \outputSet, p_j')\}_{j \in J}\) such that 
\[
    p_j' = \sum_{\substack{i \in I : \\ \oupt_j = \oupt_i \restriction_S}} p_i,
\]
and is denoted by \(\outcome((G,\localVar))[S]\) or \(\{(\oupt_i, p_i)\}_{i \in I}[S]\).
We also say that two labeling distributions \(\{(\oupt_i: V(G) \to \alphabet, p_i)\}_{i \in I}, \{(\oupt_j: V(H) \to \alphabet, p_j)\}_{j \in J}\) over two graphs \(G,H\) are isomorphic if there is an isomorphism \(\varphi: V(G) \to V(H)\) between \(G\) and \(H\) such that \(\{(\oupt_i: V(G) \to \alphabet, p_i)\}_{i \in I} = \{(\oupt_j \circ \varphi: V(G) \to \alphabet, p_j)\}_{j \in I}\).

We are now ready to define \emph{\nonsign outcomes}.
\begin{definition}[Non-signaling outcome]\label{def:non-signaling:non-signaling-outcome}
    Let \(\outcome\) be any outcome.
    Fix any two graphs \(G,H\) of \(n\) nodes and any two input data functions \(\localVar_G,\localVar_H\).
    Suppose there exists a non-negative integer \(T \ge 0\) with the following property:
    For any two subsets \(S_G \subseteq V(G), S_H \subseteq V(H)\) such that \(\varphi : V(G) \to V(H)\) is an isomorphism between \(\view_T(S_G)\) and \(\view_T(S_H)\), then the restrictions \(\outcome((H, \localVar_H))[S_H]\) and \(\outcome((G, \localVar_G))[S_G]\) are isomorphic under \(\varphi\).
    We say that \(\outcome\) is \emph{\nonsign beyond distance \(T\)} or, alternatively, that \(\outcome\) has locality \(T\).
\end{definition}

With this notion, we can define the \nonsign model.

\paragraph{The \nonsign model.} The \nonsign model is a computational model which produces \nonsign outcomes.
More specifically, the distributed network in input \((G, \localVar)\) is as in the definition of the deterministic \local model, with \(\localVar(v)\) encoding the degree of a node, port numbers, the identifier  and (possibly) an input label expected by the problem of interest.
The model can produce non-signaling outcomes where one wants to minimize the locality \(T\) to solve the problem.
Usually, we require that the success probability of an outcome that solves a problem is at least \(1 - 1/\poly(n)\). 

The \nonsign model was first introduced by \textcite{gavoille2009} with a slightly different definition, and then formalized by \textcite{arfaoui2014} in the current form.
It is stronger than any \emph{physical} synchronous distributed computing model, as \(T\)-rounds synchronous distributed algorithms (both classical and quantum) obey the no-signaling principle and must produce outcomes that are non-signaling beyond distance \(T\).
\cite{gavoille2009} was the first to observe that lower bounds in the \nonsign model must hold in all weaker models, and noticed that lower bound techniques based on the indistinguishability argument \emph{withing} the input graph family still hold in \nonsign. 
More specifically, \cite{gavoille2009} revisited some previous lower bound results and noticed that they hold also in the \nonsign model, and it also established a new lower bound for \(2\)-coloring even cycles, revisiting the first argument on indistinguishability shown in \cref{sec:non-signaling:warm-up}.

\begin{theorem}[\textcite{gavoille2009}]
    In the \nonsign model, the following holds:
    \begin{enumerate}
        \item Maximal independent requires locality \(\myOmega{\sqrt{\frac{\log n}{\log \log n}}}\) \cite{kuhn2004}.
        \item Locally minimal (greedy) coloring requires locality \(\myOmega{\frac{\log n}{\log \log n}}\) \cite{gavoille2007,gavoille2009coloring}.
        \item Finding a connected subgraph with \(\myO{n^{1 + 1/k}}\) edges requires locality \(\myOmega{k}\) \cite{derbel2008,elkin2007}.
        \item Finding a 2-coloring in even cycles requires locality \(\myOmega{\ceil{\frac{n-2}{4}}}\).
    \end{enumerate}
\end{theorem}

We remark that \cite{gavoille2009} had a slight different definition of the \nonsign model.
In the paper, the model was called \philocal and defined outcomes to exist only on a specific input graph family.
When considering, e.g., the problem of 2-coloring even cycles, the outcome produced by \philocal were defined \emph{only} for even cycles, leaving out the possibility to play with graphs outside the input graph family.
However, as previously mentioned, (classical or quantum) algorithms can be run also on network graphs that lie outside the input graph family: If computation is at any point undefined, we can let a node output some garbage label.
This freedom that we have with algorithms is what we exploit in the second lower bound argument in \cref{sec:non-signaling:warm-up}, the graph-existential one.
By defining outcomes on \emph{every possible input graph}, we have access to new lower bound techniques.
In \cite{coiteuxroy2023}, we proved that the graph-existential argument can be extended all the way up to the \nonsign model (under some general hypotheses), and used it to prove the following lower bounds.

\begin{theorem}[\textcite{coiteuxroy2023}]\label{thm:non-signaling:new-lbs}
    In the \nonsign model, the following holds:
    \begin{enumerate}
        \item The problem of \(c\)-coloring \(\chi\)-chromatic graphs requires locality \(\myOmega{n^{1/{\floor{\frac{c-1}{\chi - 1}}}}}\).
        \item Finding a \(3\)-coloring \(n \times m\) grids requires locality \(\min\{n,m\}\).
        \item Finding a \(c\)-coloring of trees requires locality \(\log_c n\).
    \end{enumerate}
\end{theorem}
The three results of \cref{thm:non-signaling:new-lbs} make use of \emph{cheating graphs}, that is, of graphs that are locally everywhere indistinguishable from proper inputs, but globally they are not in the input family.
E.g., for \(c\)-coloring \(\chi\)-chromatic graphs we need to find a graph such that the radius-\(T\) neighborhood of any node induces a graph that is \(\chi\)-colorable, for \(T = \myTheta{n^{1/{\floor{\frac{c-1}{\chi - 1}}}}}\), but globally the graph has chromatic number strictly greater than \(c\).
The existence of such graph immediately implies that deterministic \local algorithms require locality \(\myOmega{n^{1/{\floor{\frac{c-1}{\chi - 1}}}}}\) to solve the problem.
Such graph is given, for all combinations of \(c\) and \(\chi\), by \textcite{bogdanov2013}: 
this was the first time that \citeauthor{bogdanov2013}'s construction found application in distributed computing and, more in general, theoretical computer science.
Interestingly, in \cite{coiteuxroy2023} the authors proved an almost matching upper bound for the problem in \detlocal, thus excluding any significant quantum advantage over \detlocal for this problem.
As for \(3\)-coloring grids, we made use of odd quadrangulations of Klein-bottles \cite{mohar2002}, which are everywhere locally indistinguishable from grids, but have global chromatic number, while for \(c\)-coloring trees we revisited \citeauthor{linial92}'s argument in \cite{linial92} that made use of Ramanujan graphs, which are high-girth, high-chromatic graphs \cite{mohar2013}.
When looking at \randlocal and the \nonsign model, we need to make sure that the graph can be nicely covered by a small amount of subgraphs (whose union form the whole graph) that are slightly overlapping, in order to identify a subgraph where the failure probability is high enough.
In general, boosting the failure probability is possible also in the \nonsign model. 
However, we cannot rely on the same argument of \cref{sec:non-signaling:warm-up} as a non-signaling outcome does not guarantee \emph{independence} between the output distributions of far-away subsets of nodes (which instead is guaranteed by \randlocal algorithms).
The reason is that \nonsign outcomes include the possible use of \emph{global} resources, such as \emph{shared randomness} or \emph{shared quantum states}.
Such resources make output distributions of distant nodes dependent of each other, even if their distance is greater than the locality of the algorithm itself.
Nonetheless, boosting the failure probability is still possible provided that the cheating graph meets some properties.
For more details, we defer the reader to the original article \cite{coiteuxroy2023}.

All this discussion might suggest that \nonsign argument are sufficient to exclude quantum advantage, and hence gives rise to the following question:
\begin{question}
    \label{question:non-signaling:avoiding-advantage}
    Can we exclude quantum advantage for all LCL problems using \nonsign arguments?
\end{question}
Unfortunately, the answer to this question is \emph{no}. 
Indeed, the \nonsign model is too strong to compare with classical \local.
For example, very recently \textcite{balliu2024shared-randomness} solved a longstanding open problem, proving that shared randomness gives advantage over private randomness in classical \local when restricting to LCL problems (otherwise, the thesis is trivial).  
More specifically, there is an LCL problem that requires \(\myOmega{\sqrt{n}}\) rounds in \randlocal but can be solved in \(\myO{\log n}\) rounds when nodes have access to shared randomness (e.g., an infinite random bit string).
Since the \nonsign model is strong enough to simulate \local with shared randomness, we already know that it is too strong with respect to classical \local and there is no hope to prove something like \cref{question:non-signaling:avoiding-advantage}.
However, it is worth investigating to what extent lower bound arguments based on the \nonsign property apply, since we still miss a characterization.
\begin{question}[Open]\label{question:non-signaling:to-what-extent}
    For which LCL problems can we exclude quantum advantage using \nonsign arguments?
\end{question} %
\section{The \boundep model}
\label{sec:bounded-dependence}

\begin{figure}[h!]
    \centering
    \includegraphics[scale=0.6]{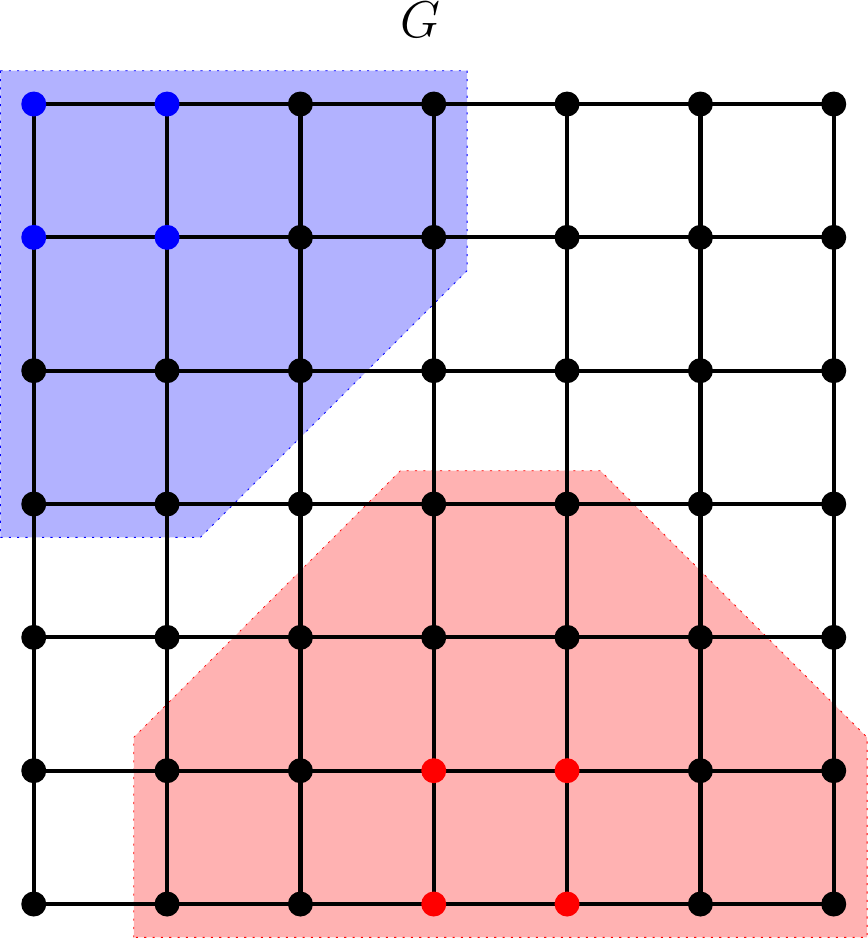}
    \caption{Bounded-dependence property.
    When running any \(2\)-round synchronous distributed algorithm (which does not rely on shared resources), the output labeling distribution of the red nodes is independent of that of the blue nodes.}
    \label{fig:bounded-dependence:property}
\end{figure}

To avoid dealing with shared resources, we can \emph{weaken} the \nonsign model by introducing further restrictions on the \nonsign outcomes.
Running \(T\)-rounds classical or \qlocal algorithms without shared resources, we obtain the following property on the output labeling distribution: for every two subset of nodes \(A,B\) such that \(\dist(A,B) > 2T\), then the output distributions restricted to \(A\) and \(B\) are independent (see \cref{fig:bounded-dependence:property}).
Let us formalize this notion.

\begin{definition}[\(T\)-dependent distribution]
    Let \(\outputSet\) and \(I\) be two sets, and let \((G,\localVar)\) be an input for some labeling problem \(\problem\).
    An output labeling distribution \(\{(\oupt_i : V(G) \to \outputSet, p_i)\}_{i \in I}\) is  \(T\)-dependent if, for any two subsets of nodes \(A,B \subseteq V(G)\) such that \(\dist(A,B) > T\), we have that \(\{(\oupt_i, p_i)\}_{i \in I}[A]\) is independent of \(\{(\oupt_i : V(G) \to \outputSet, p_i)\}_{i \in I}[B]\).
\end{definition}

We can think now of \nonsign outcomes that produces such distributions.

\begin{definition}[Bounded-dependent outcome]\label{def:bounded-dependence:bounded-dependent-outcome}
    Let \(\outcome\) be any outcome that is \nonsign beyond distance \(T\).
    We say that \(\outcome\) is \boundept with locality \(T\) if for any input \((G, \localVar)\) we have that \(\outcome((G,\localVar))\) is \(2T\)-dependent.
    Furthermore, when \(T = \myO{1}\) (i.e., it does not depend on the input graph), we say that \(\outcome((G,\localVar))\) is a finitely-dependent distribution.
    Moreover, if for all inputs \((G,\localVar)\) it holds that \(\outcome((G,\localVar))\) does not depend on identifiers and port numbers, we say that \(\outcome\) is \emph{invariant under subgraph isomorphism}.
\end{definition}

With the addition of this further property, we can define the \emph{\boundep model}, first formalized in \cite{akbari2024}.

\paragraph{The \boundep model.}
Similarly to the introduction of the \nonsign model, we can define the \emph{\boundep model} as a model that produces bounded-dependent outcomes.
Again, the required success probability then solving a problem should be at least \(1 - 1/\poly(n)\).

One might hope that we can prove lower bounds in the \boundep model and matching upper bounds in classical \local.
\begin{question}\label{question:bounded-dependence:advantage}
    Can we always rule out quantum advantage for LCLs using bounded-dependent outcomes?
\end{question}
Unfortunately, the answer to \cref{question:bounded-dependence:advantage} is \emph{no}: \textcite{holroyd2016,holroyd2018} proved it for us.
It is well-known that \(3\)-coloring paths and cycles requires locality \(\myTheta{\log^\star n}\) both in \detlocal and in \randlocal \cite{linial92,chang16exponential}. 
\cite{holroyd2016,holroyd2018} showed that these problems admit finitely-dependent distributions, that is, it is possible to \(3\)-color path and cycles with an \(\myO{1}\)-dependent outcome that is invariant under subgraph isomorphism and under permutations of the colors. 

Apart from \(3\)-coloring paths and cycles, there are other famous problems in the complexity class \(\myTheta{\log^\star n}\), such as computing an MIS, \((\maxDeg + 1)\)-coloring graphs of maximum degree \(\maxDeg\), etc.

All such problems are also called \emph{symmetry-breaking} problems, in the sense that they are easily solvable by an \(\myO{1}\)-round \portnum algorithm (and without knowledge of \(n\)) if symmetry is locally broken with a constant amount of labels.
This fact implies that, given any two LCLs \(\problem_1,\problem_2\) that have complexity \(\myTheta{\log^\star n}\) in some graph \(G\), a solution to any of those problems can be converted into a solution of the other in a constant number of rounds.
In \cite{akbari2024}, the authors proved it is possible to compose \portnum algorithms (that do not make use of \(n\)) and \boundept outcomes obtaining a \boundept outcome (without significant loss in locality).
Hence, the results in \cite{holroyd2016,holroyd2018} immediately imply that in paths and cycles all LCL problems that have classical complexity \(\myTheta{\log^\star n}\) are solvable by an \(\myO{1}\)-dependent outcome in the \boundep model as well: but what about other graphs?

\begin{question}\label{question:bounded-dependence:log-star}
    Is there any LCL problem with classical complexity \(\myTheta{\log^\star n}\) for which we can rule out quantum advantage using the \boundep model?
\end{question}

Unfortunately the answer is, again, \emph{no}.
\textcite{akbari2024} proved that all such problems are solvable with locality \(\myO{1}\) in the \boundep model, and the resulting \boundept outcomes are also invariant under subgraph isomorphism.

\begin{theorem}[\textcite{akbari2024}]\label{thm:bounded-dependence:log-star}
    Let \(\problem\) be an LCL over some input graph family \(\FF\) that has complexity \(\myO{\log^\star n}\) in classical \local.
    Then, there exists an \(\myO{1}\)-dependent outcome \(\outcome\) that solves \(\problem\) over \(\FF\).
    Furthermore, \(\outcome\) is invariant under subgraph isomorphism.
\end{theorem}

\cref{thm:bounded-dependence:log-star} is powerful result that shows the power of finitely-dependent distributions: 
such distributions are able to break symmetry with constant locality, that is something that classical \local cannot do.
This leaves us with one of the major open question in the field.
\begin{question}[Open]\label{question:bounded-dependence:major1}
    Is \qlocal able to break symmetry in LCLs? That is, can \qlocal solve in time \(\mylittleo{\log^\star n}\) any LCL \(\problem\) that has classical complexity \(\myTheta{\log^\star n}\)?
\end{question}
We currently lack tools to analyze directly \qlocal (especially regarding lower bounds).
What we can do is instead focusing on specific graph families and/or specific complexity classes in the \boundep or \nonsign model.
In classical (both deterministic and randomized) \local, LCL complexities belong to the following three classes: \(\myO{1}\), \(\myTheta{\log^\star n}\), and \(\myOmega{\log \log n}\).
After \cref{thm:bounded-dependence:log-star}, we might wonder what happens in the complexity class \(\myOmega{\log \log n}\).

\begin{question}[Open]\label{question:bounded-dependence:major2}
    In the \boundep model and the \nonsign, what can we infer on the complexity of LCL problems that have classical complexity \(\myOmega{\log \log n}\)?
\end{question}

One of the most prominent problems is maybe sinkless orientation (SO), a problem that asks each node \(v\) to orient its adjacent edges so that \(\outdeg(v) \ge 1\).
It is known that SO has complexity \(\myTheta{\log n}\) in \detlocal and \(\myTheta{\log \log n}\) in \randlocal \cite{balliu22so-simple}.
Hence, we can formulate the following question, which is nowadays open.

\begin{question}[Open]\label{question:bounded-dependence:sinkless-orientation}
    What is the complexity of sinkless orientation in the \boundep model or the \nonsign model?
\end{question}

Currently, we have little insight on these questions, but we managed to give some small partial answers to \cref{question:bounded-dependence:major2}, which we address in the following section. %
\section{The \olocal model}

In this section, we describe other models of computation that, at a first glance, seem completely unrelated from the (super)quantum world we have been describing so far, but after a deeper look turn out to be related and extremely useful.
In \cite{ghaffari2017}, \citeauthor{ghaffari2017} introduced the \slocal model, that is, a sequential version of the \local model.

\paragraph{The \slocal model.}
The \slocal model is similar to the \local model, but sequential.
Here, the nodes of the input graph \(G = (V,E)\) with \(\abs{V} = n\) are processed according to an adversarial order \(\sigma = v_1, \dots, v_n\).
While processing a node \(v_i\), a \(T\)-round algorithm collects all data contained in and the topology of the radius-\(T\) neighborhood of \(v_i\) (including the states and the outputs of previously processed nodes in \(\neighborhood_T[v_i]\), i.e., any node \(v_j \in \neighborhood_T[v_i]\) for \(j < i\)): 
We say that such an algorithm has locality/complexity/running time \(T\).
Note that the algorithm might store the whole data contained in \(\view_T(v_i)\) inside the memory of \(v_i\) (and we always assume this happens, since it can only make the algorithm stronger).
This phenomenon gives to a node \(v_i\) access to the data contained in  \(\view_T(v_j)\) if and only if there is a subsequence of nodes \(\{v_{h_k}\}_{k \in [m]}\) with \(j = h_k < h_{k+1} < \dots < h_m = i\) such that \(v_{h_k} \in \NN_T[v_{h_{k+1}}]\) for all \(k \in [m-1]\).

If the algorithm is given in input an infinite random bit string, we talk about the randomized \slocal model, as opposed to the deterministic \slocal model. 
Notice that the adversarial order in which node are processed is assumed to be \emph{oblivious} to the random bit string.
In this case, we require the success probability to be at least \(1 - 1/\poly(n)\), with \(n\) being the number of nodes of the input graph.

Clearly, the \slocal model is stronger than the \local model, since any deterministic \(T\)-round \local algorithm can be converted into a deterministic \(T\)-round \slocal algorithm.
Surprisingly, \textcite{ghaffari2018derandomizing} proved that also any randomized \(T\)-round \local algorithm can be converted into a deterministic \(\myO{T}\)-round \slocal algorithm through some derandomization technique.
Interestingly, \cite{ghaffari2017} proved that, under certain hypotheses, (both randomized and deterministic) \slocal algorithms can be converted in (respectively, randomized and deterministic) \local algorithms with some overhead in the complexity.
However, for our purposes it is sufficient to know that \(\myO{1}\)-round \slocal algorithms for LCLs can be turned into \(\myO{\log^\star n}\)-round \local algorithms: this is folklore, but a proof can be found in \cite{akbari2024}.
See \cref{fig:landscape-of-models} for a representation of the relations among models.

On top of the \slocal model, \textcite{akbari_et_al:LIPIcs.ICALP.2023.10} introduced the \olocal model. 
The \olocal model is simply the \slocal model equipped with global memory.

\paragraph{The \olocal model.}
The (deterministic) \olocal model is basically equivalent to the \slocal model with global memory. 
More specifically, the \olocal model is a centralized model of computing where the algorithm initially knows only the set of nodes of the input graph $G$. 
The nodes are processed with respect to an adversarial input sequence $\sigma = v_{1}, v_{2}, \dots, v_{n}$.
The output of $v_{i}$ depends on $G_{i} = \cup_{j=1}^i \view_T(v_{j})$, i.e., the radius-\(T\) views of of $v_{1}, v_{2}, \dots, v_{i}$ (which includes all input data), plus all the outputs of previously processed nodes (we can imagine that the views get updated at each step of the algorithm).

In \cite{akbari2024}, the authors defined the randomized \olocal model as a randomized variant of the \olocal model where the label assigned by the algorithm to $v_{i}$ might depend on arbitrarily large portions of an infinite random bit string. 
Note that this model is oblivious to the randomness used by the algorithm. 
In particular this means that the graph outside $G_i$ cannot be changed depending on the label assigned to $v_{i}$. 
One could also define the randomized \olocal model in an adaptive manner, but it turns out that this is equivalent to the deterministic \olocal model (as proved in \cite{akbari2024}).
The notion of complexity of a problem can be easily extended from the \local model.
If the algorithm is randomized, we also require that the failure probability is at most \(1/\poly(n)\), where \(n\) is the size of the input graph.

Interestingly, \cite{akbari_et_al:LIPIcs.ICALP.2023.10} proved that in rooted regular trees LCLs have roughly the same complexity across the three deterministic models \local, \slocal, and \olocal.

So, why should we care about the \olocal model?
\textcite{akbari2024} found a very important connection with the \nonsign model, which we report here with the following theorem: in \cite{akbari2024} the result is stated only for LCL problems, but the proof holds also for any labeling problem.

\begin{theorem}[\textcite{akbari2024}]\label{thm:online-local:non-signaling-simulation}
    Let \(\problem\) be any labeling problem and \(\outcome\) any \nonsign outcome with locality \(T\) solving \(\problem\) with probability \(p > 0\).
    Then, there is a randomized \olocal algorithm \(\algo\) with locality \(T\) that solves \(\problem\) with probability \(p\).
    Furthermore, the output distribution of \(\algo\) over any input \((G,\localVar)\) is exactly \(\outcome((G,\localVar))\).
\end{theorem}

\cref{thm:online-local:non-signaling-simulation} is very powerful, as it allows us to focus on classical, centralized models of computing: every lower bound in randomized \olocal holds also in \nonsign and, hence, in \qlocal.
Surprisingly, \cite{akbari2024} proved even more results regarding the \olocal model that connects it with the classical \local model.

In order to introduce it, we need to define \emph{component-wise} \olocal algorithms.
\paragraph{The component-wise \olocal model.}
The component-wise \olocal model is exactly the same as the deterministic \olocal model but when the algorithm processes a node \(v_i\) according to the adversarial order \(\sigma = v_1, \dots, v_n\), \(v_i\) does not have access to the whole \(G_i\).
Rather, it has access only to its connected component in \(G_i\).
Clearly a component-wise \olocal algorithm is also a standard \olocal algorithm, hence the model is weaker than the deterministic \olocal model  (see \cref{fig:landscape-of-models} for a landscape of all the computational models).
However, we have the following result.

\begin{theorem}[\textcite{akbari2024}]\label{thm:obline-local:component-wise-reduction}
    Let \(\problem\) be any LCL problem, and let \(\algo\) be any \olocal algorithm solving \(\problem\) with locality \(T(n)\) for graphs of \(n\) nodes.
    Then the following holds:
    \begin{enumerate}
        \item If \(\algo\) is deterministic, then there exists a deterministic component-wise \olocal algorithm solving \(\problem\) with locality \(T(2^{O({n^3})})\).
        \item If \(\algo\) is randomized and has success probability \(p(n) > 0\), then there exists a deterministic component-wise \olocal algorithm solving \(\problem\) with locality \(T\left(2^{O({n^3})} + 2^{O(2^{n^2})}\cdot \log \frac{1}{p(n)}\right)\).
    \end{enumerate}
\end{theorem}

\subsection{Implications in rooted trees}

Why is \cref{thm:obline-local:component-wise-reduction} so meaningful?
The difference between \olocal and \slocal is \emph{local} vs \emph{global memory}.
Component-wise \olocal algorithms lie somewhere in the middle: a node gets access only to the memory contained in the currently explored connected component.
Interestingly, in some topologies, the \slocal model is able to ``simulate'' component-wise \olocal algorithms.
We remind the reader that a rooted tree is a directed tree where all nodes have outdegree \(1\) except for a single node \(v\) that has outdegree \(0\) and is called the \emph{root} of the tree.

\begin{theorem}[\textcite{akbari2024}]\label{thm:online-local:component-wise-to-slocal}
    Let \(\problem\) be any LCL problem over rooted trees. 
    Assume \(\algo\) is a component-wise \olocal algorithm solving \(\problem\) with locality \(T(n)\).
    Then, there exists a deterministic \slocal algorithm solving \(\problem\) with locality \(\myO{1} + T(\myO{n})\).
\end{theorem}

Now we can go from (deterministic or randomized) \olocal to \slocal in rooted trees.
Interestingly, \cite{akbari2024} also proved that LCLs over rooted trees in \slocal have complexity either \(\myO{1}\) or \(\myOmega{\log n}\).

\begin{theorem}[\textcite{akbari2024}]\label{thm:online-local:slocal-rooted-trees-constant}
    Let \(\problem\) be any LCL problem over rooted trees.
    Assume \(\algo\) is an \slocal algorithm solving \(\problem\) with locality  \(\mylittleo{\log n}\).
    Then, there exists another \slocal algorithm \(\BB\) solving \(\problem\) with locality \(\myO{1}\).
\end{theorem}

It is folklore that LCLs that have complexity \(\myO{1}\) in \slocal over any topology translate in complexity \(\myO{\log^\star n}\) in classical \local. 
Altogether, we have the following.

\begin{corollary}[\textcite{akbari2024}]\label{cor:online-local:olocal-to-local-rooted-trees}
    Let \(\problem\) be any LCL problem over rooted trees.
    Then the following holds:
    \begin{enumerate}
        \item If \(\problem\) has complexity \(\mylittleo{\log \log n}\) in deterministic \olocal, then it has complexity \(\myO{\log^\star n}\) in \local.
        \item If \(\problem\) has complexity \(\mylittleo{\log \log \log n}\) in randomized \olocal, then it has complexity \(\myO{\log^\star n}\) in \local.
    \end{enumerate}
\end{corollary}

See also \cref{fig:landscape-of-models} for a drawing of all the implications among models.
\cref{cor:online-local:olocal-to-local-rooted-trees} is very powerful in a twofold sense.
On one hand, it implies that the LCL complexity class \(\myO{\log^\star n}\) in \qlocal and in \local coincide over rooted trees, excluding significant quantum advantage (it might still be that \(\myO{1}\) locality in \qlocal becomes \(\myTheta{\log^\star n}\) in \local). 
Also, the LCL complexity class \(\myO{1}\) in the \boundep and \nonsign models becomes \(\myO{\log^\star n}\) in \local over rooted trees, while we know that \(\myO{\log^\star n}\) in \local becomes \(\myO{1}\) in the \boundep and \nonsign models over any topology by \cref{thm:bounded-dependence:log-star}.
On the other hand, \cref{cor:online-local:olocal-to-local-rooted-trees} allows us to obtain lower bounds in \olocal (and, hence, \qlocal, \boundep, and \nonsign) through lower bounds in \local: 
We know that if an LCL over rooted trees cannot be solved in time \(\myO{\log^\star n}\) in \local, then it needs locality \(\myOmega{\log \log n}\) in deterministic \olocal and \(\myOmega{\log \log \log n}\) in randomized \olocal.
Hence, in rooted trees we also know that the LCL complexity class \(\mylittleomega{\log^\star n}\) in \local becomes \(\myOmega{\log \log \log n}\) in \qlocal, \boundep, and \nonsign.

\subsection{LCL complexity landscape in general trees}

Recently, \textcite{dhar24rand} analyzed more carefully the relation between the randomized \olocal model and the \detlocal model in various kind of trees: rooted, unrooted, regular, etc.
They proved results for the LCL complexity class \(\mylittleomega{\log n}\) by extending previous results all the way up to the randomized \olocal model \cite{akbari_et_al:LIPIcs.ICALP.2023.10,balliu22regular-trees,balliu22rooted-trees,balliu20almost-global,chang20,chang19hierarchy,grunau2022}.
By these previous works, it was known that in the \local model, LCL problems over regular trees that have complexity \(\mylittleomega{\log n}\) fall into one of the following classes: \(\myTheta{n^{1/k}}\) for some \(k \in \natsPos\).
Furthermore, if the tree is also rooted, we know that the only possible complexities in \detlocal are \(\myO{1}\), \(\myTheta{\log^\star n}\), and \(\myTheta{\log n}\).
In \cite{akbari_et_al:LIPIcs.ICALP.2023.10} this result was extended all the way up to deterministic \olocal (but only for the case of rooted trees, and did not find an \emph{exact} correspondence between \local and \olocal).
The authors of \cite{dhar24rand} proved the following two theorems.

\begin{theorem}[\textcite{dhar24rand}]
    Let \(\problem\) be an LCL problem on unrooted regular trees.
    If \(\problem\) has complexity \(\myTheta{n^{1/k}}\) for any \(k \in \natsPos\) in \detlocal, then it has  complexity \(\myTheta{n^{1/k}}\)  in randomized \olocal, and vice-versa.
\end{theorem}

\begin{theorem}[\textcite{dhar24rand}]\label{thm:online-local:rooted-regular-trees}
    Let \(\problem\) be an LCL problem on rooted regular trees.
    The following holds:
    \begin{enumerate}
        \item If \(\problem\) has complexity \(\myTheta{n^{1/k}}\) for any \(k \in \natsPos\) in \detlocal, then it has  complexity \(\myTheta{n^{1/k}}\)  in randomized \olocal, and vice-versa.
        \item If \(\problem\) has complexity \(\myTheta{\log n}\) in \detlocal, then it has  complexity \(\myTheta{\log n}\)  in randomized \olocal, and vice-versa.
    \end{enumerate}
\end{theorem}

Combining \cref{thm:online-local:rooted-regular-trees} with \cref{cor:online-local:olocal-to-local-rooted-trees}, we obtain that in the LCL complexity region \(\myO{\log n}\) over rooted regular trees, the following classes contain the same problems in \local and randomized \olocal, and are the only possible complexity classes: \(\myO{\log^\star n}\) in \detlocal and \(\myO{1}\) in randomized \olocal, or \(\myTheta{\log n}\) both in \detlocal and randomized \olocal.
We remind the reader that all these equivalences between complexity classes also hold between \local and \qlocal, as well as the \boundep and the \nonsign models, as they are ``sandwiched'' between \detlocal and randomized \olocal.
\cite{dhar24rand} also provided a general result on trees extending speedup arguments in \cite{balliu20almost-global}.

\begin{theorem}[\textcite{dhar24rand}]\label{thm:online-local:general-trees}
    Let \(\problem\) be an LCL problem on general trees.
    Then, either \(\problem\) has complexity \(\myTheta{n}\) in the (deterministic or randomized) \local model and in the (deterministic or randomized, respectively) \olocal model, or the complexity in both models is \(\myO{\sqrt{n}}\).
\end{theorem}

\begin{figure}[h!]
    \centering
    \includegraphics[scale=0.8]{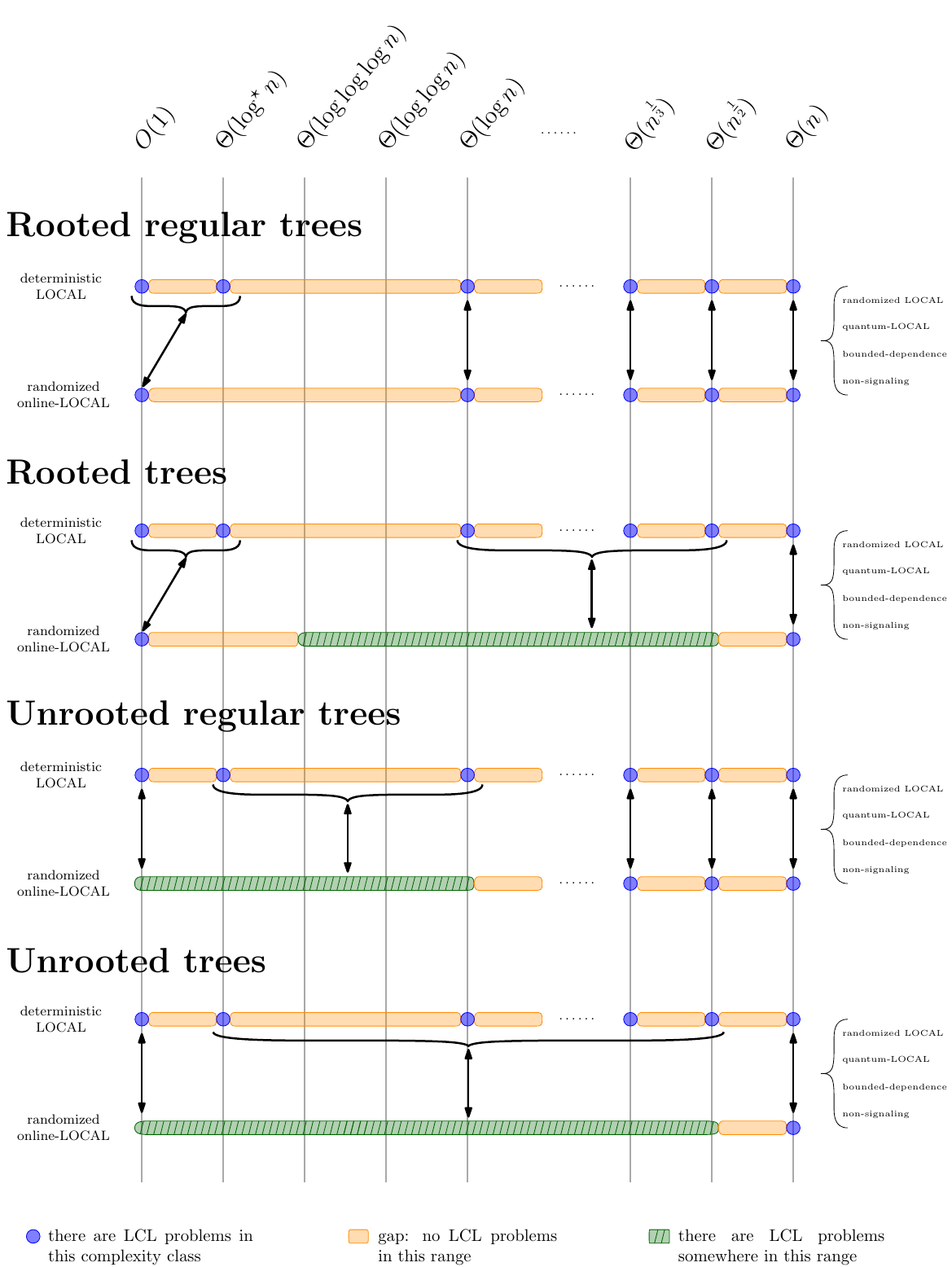}
    \caption{Landscape of LCL complexities in trees.}
    \label{fig:online-local:landscape-trees}
\end{figure}

\cref{thm:online-local:general-trees} trivially holds for any intermediate model instead of randomized \olocal.
We invite the reader to have a look at \cref{fig:online-local:landscape-trees} for a representation of the LCL complexity landscape in trees after the results of \cite{akbari2024,dhar24rand}.\footnote{We reproduced Figure 2  in \cite{dhar24rand}, for which we give credits to the authors.}
We conclude the section observing that, perhaps, the most difficult complexity range is that between \(\mylittleomega{\log^\star n}\) (hence, by known results, \(\myOmega{\log \log n}\)) and \(\myO{\log n}\) in general (regular or non-regular) trees.
The reason is that we don't have techniques that we can refer to which we can extend to these (super)quantum models.

\begin{question}[Open]\label{question:online-local:open}
    Let \(\problem\) be an LCL problem over (regular or non-regular) trees that has classical complexity  between \(\myOmega{\log \log n}\) and \(\myO{\log n}\).
    Can we find non-trivial upper or lower bounds on the complexity of \(\problem\) in randomized \olocal?
\end{question}

Again, sinkless orientation is an example of such problem.
The lower bound of sinkless orientation (and often of other LCLs with similar complexity) is proved in classical \local via a famous lower bound technique known as round elimination \cite{brandt16lll,Brandt2019automatic}.
As we argue in the next section, round elimination is unfortunately not generalizable to (super)quantum models.
Other candidate problems are: 3-coloring trees, 2-2-3 in 3-regular trees (that is, \(2\)-coloring trees so that each node of any color \(x\) has at least \(2\) neighbors colored with a color that is different from \(x\)), \(\maxDeg\)-coloring trees of maximum degree \(\maxDeg\), etc. %
\section{Quantum advantage for a local problem}

What do we know about quantum advantage in the \local model?
Before last year, there was only one example of quantum advantage: \textcite{legall2019} showed that there exists a problem that requires \(\myOmega{n}\) communication rounds in input graphs with \(n\) nodes in classical \local, but can be solved in \(\myO{1}\) rounds in \qlocal.
This problem, however, has an inherent global and artificial definition, and the winning condition depends on the joint output of nodes that are at distances \(\myOmega{n}\) between one another, which makes it very far from problems that are usually interesting for the distributed computing community, especially from LCLs.
Last year, \textcite{balliu2024quantum} exhibited the first \emph{local problem} \(\problem\) that admits quantum advantage in the \local model.
The authors proved that \(\problem\) is solvable in \(\myO{1}\) rounds in \qlocal, but requires \(\myTheta{\maxDeg}\) rounds in classical \local, where \(\maxDeg\) is the degree of the input graph.
Before describing the problem in details, let us remark that \(\problem\) is locally checkable in all senses except that the maximum degree of the input graph is not bounded, hence it is not an LCL in the strict sense of \cref{def:preliminaries:lcls}.
Indeed, in case \(\maxDeg = \myO{1}\), then there would be no asymptotic difference between the complexities in \local and \qlocal.
Let us now introduce the problem in multiple steps: first we introduce the GHZ game between three players, then we use it to build \(\problem\).

\subsection{The GHZ game}\label{sec:quantum-advantage:GHZ}
The GHZ game is a game between three players that works as follows: 
Alice, Bob, and Charlie all receive by an adversary one input bit.
The input \((x,y,z)\) is drawn from the set \(\{(0,0,0),(0,1,1),(1,0,1),(1,1,0)\}\), that is, the promise is that \(x \oplus y \oplus z = 0\) (\(\oplus\) is the notation for the XOR logical operator).
The players must produce one output bit each, resulting in a tuple \((a,b,c)\) such that \(a \oplus b \oplus c = 0\) if and only if \((x,y,z) = (0,0,0)\) (see \cref{tab:quantum-advantage:ghz_game}).
Now, the game allows players to agree on a strategy \emph{before} receiving the input  \((x,y,z)\), but \emph{after} it they cannot communicate anymore.
The best classical strategy for the three players to win this game is to deterministically output a tuple that wins in case \((x,y,z) \neq (0,0,0)\).
However, in the quantum world there is a strategy that \emph{always} wins the game.
An uninterested reader might just skip the rest of this subsection, as the only notion that is useful for the rest of this article is that there exists a quantum strategy that always wins the game.
Otherwise, we assume the reader is familiar with basic notions of quantum computing: if not, we defer the reader to introductory surveys like, e.g., \cite{BennettS98}.
The players can share a tripartite entangled state \(\ket{\psi} = \frac{1}{\sqrt{2}}(\ket{000} + \ket{111})\), known as the GHZ state.
If a player receives \(0\) as input, it makes a measurement of the entangled state in the basis \(\{\ket{+},\ket{-}\}\).
Otherwise, it makes a measurement in the basis \(\{\frac{1}{\sqrt{2}}(\ket{0} + i \ket{1}), \frac{1}{\sqrt{2}}(\ket{0} - i \ket{1})\}\).
In both cases, the player outputs \(0\) if the result of the measurement is the first state, and \(1\) if it is the second state \cite{Brassard2004QuantumP}.

\begin{table}[h!]
    \centering
    \renewcommand{\arraystretch}{1.5}
    \begin{tabular}{|c|c|c|c|}
        \hline
        \textbf{Alice's Input (\(x\))} & \textbf{Bob's Input (\(y\))} & \textbf{Charlie's Input (\(z\))} & \textbf{Winning Condition (Outputs)} \\ \hline
        0 & 0 & 0 & \(a \oplus b \oplus c = 0\) \\ \hline
        0 & 1 & 1 & \(a \oplus b \oplus c = 1\) \\ \hline
        1 & 0 & 1 & \(a \oplus b \oplus c = 1\) \\ \hline
        1 & 1 & 0 & \(a \oplus b \oplus c = 1\) \\ \hline
    \end{tabular}
    \caption{GHZ game definition: the tuple \((x,y,z)\) is the input and the output \((a,b,c)\) must satisfy the winning condition.}
    \label{tab:quantum-advantage:ghz_game}
\end{table}

\subsection{Iterated GHZ games}\label{sec:quantum-advantage:iteratedGHZ}
The GHZ game usually comes with a promise that \(x \oplus y \oplus z = 0\).
In order to define our labeling problem, we need to remove this promise.
We do so by relaxing the definition of the game: when \(x \oplus y \oplus z = 1\) we allow any combination of outputs from Alice, Bob, and Charlie: clearly,this modification cannot make the problem harder.
Now, our input is a bipartite graph where the two sets of the bipartition are the set \(W\) of \emph{white nodes} and the set \(B\) of \emph{black nodes}.
White nodes are the real players of the games, while black nodes represent the games that the players are playing.
More specifically, we assume that each white node has degree \(\maxDeg\) and each black node has degree \(3\).
We also assume that black nodes are colored with colors from \([\maxDeg]\), and that each white node has a unique neighbor colored with color \(c\) for all \(c\in [\maxDeg]\).
The colors of the black nodes specify the order according to which a white node must play the games.
Every white node \(v\) must output a vector \([v_1, \dots, v_\maxDeg]\) of \(\maxDeg\) bits, where \(v_i \in \{0,1\}\) is its output to the game represented by its neighboring black node of color \(i\).
Let \(v\) be any black node colored with color \(c \ge 1\), and \(s,t,u\) its three neighboring white nodes.
The problem is defined as follows:
\begin{enumerate}
    \item If \(c = 1\), then exactly one among \(s_1\), \(t_1\), and \(u_1\) must be equal to 1, and the others must be equal to 0.
    \item If \(c > 1\), then: 
        \begin{enumerate}
            \item If \(s_{c-1} \oplus t_{c-1} \oplus u_{c-1} = 0\), then \(s_c\), \(t_c\), and \(u_c\) must solve the GHZ game with input \((s_{c-1}, t_{c-1}, u_{c-1})\), that is, \(s_{c} \oplus t_{c} \oplus u_{c} = 0\) if \(s_{c-1}, t_{c-1}, u_{c-1} = (0,0,0)\), and \(s_{c} \oplus t_{c} \oplus u_{c} = 1\) otherwise.
            \item If \(s_{c-1} \oplus t_{c-1} \oplus u_{c-1} = 1\), then \(s_c\), \(t_c\), and \(u_c\) can be arbitrarily chosen.
        \end{enumerate}
\end{enumerate}

Clearly, in 1 communication round the problem can be solved in \qlocal as follows:
The black nodes of color \(c = 1\) inform their three neighbors \(s\), \(t\), and \(u\) on how to set \(s_1\), \(t_1\), and \(u_1\) so that exactly one of these outputs is equal to \(1\).
Black nodes of color \(c > 1\) prepare \(3\) entangled GHZ states as described in \cref{sec:quantum-advantage:GHZ} and send them to their three white neighbors, along with their colors.
The white node will now have \(\maxDeg-1\) entangled states \(\ket{\psi_2}, \dots, \ket{\psi_\maxDeg}\), where \(\ket{\psi_c}\) comes from the black neighbor of color \(c\).
Then, it can locally measure the entangled states, in order, setting all correct outputs without communicating further, as output number \(i\) depends only on \(\ket{\psi_i}\) and output number \(i-1\).

In the classical setting, a trivial strategy would be the following:
In round \(1\), black nodes of color 1 inform their white neighbors about which one should set its output to 1, while the others set their output to 0. 
In round \(2\), the white nodes inform their neighbors of color \(2\) about their output that has been set in the previous round.
In this way, the three white neighbors \(s\), \(t\), and \(u\) of a black node \(v\) of color \(2\) have specified the inputs \(s_1\), \(t_1\), and \(u_1\) for another GHZ game. Now \(v\) can locally solve the GHZ game with the new inputs and inform \(s\), \(t\), and \(u\) about their new outputs \(s_2\), \(t_2\), and \(u_2\).
Iterating this procedure, we have a solution to the problem in \(2\maxDeg\) rounds.

Interestingly, \textcite{balliu2024quantum} proved the following result.

\begin{theorem}[\textcite{balliu2024quantum}]\label{thm:quantum-advantage:iteratedGHZ}
    The iterated GHZ problem can be solved in \(1\) round in \qlocal but requires \(\myOmega{\min\{\maxDeg, \log_\maxDeg \log n\}}\) rounds in classical (deterministic or randomized) \local.
\end{theorem}

The classical lower bound in \cref{thm:quantum-advantage:iteratedGHZ} is obtained through \emph{round elimination} (RE), one of the most prominent lower bound techniques.
Round elimination was formalized by \textcite{Brandt2019automatic} but already used in a less general form to get an \(\myOmega{\log^\star n}\) lower bound for \(3\)-coloring cycles by \textcite{linial92}.
Nowadays, we even have an automated software that guides us into the round elimination procedure (i.e., REtor \cite{Olivetti2019}).
Round elimination works as follows: 
Suppose an LCL problem \(\problem\) has some complexity \(T\), that is, there exists a \(T\)-round \local algorithm \(\algo\) that solves \(\problem\), and \(T\) is the minimum integer with such property.
Imagine running \(\algo\) for \(T-1\) rounds on a graph \(G\).
Nodes of \(G\) now will contain enough knowledge to be able to solve \(\problem\) with just one more round of communication.
Hence, one can describe exactly what this knowledge is and come up with the most general LCL problem \(\problem_1\) nodes can solve in \(T-1\) rounds with \(\algo\). 
Now suppose we iterate this procedure for \(T\) rounds: We end up with an LCL \(\problem_T\) that is solvable in \(0\) rounds of communication.
However, \(T\) is usually unknown and the object of our investigation. 
Hence, we can guess the magnitude of \(T\) (say, \(T'\)) and analyze \(\problem_{T'}\).
If the description of \(\problem_{T'}\) is simple enough, it might be incredibly easier to understand if \(\problem_{T'}\) is solvable in \(0\) rounds: 
if not, then we get a lower bound \(T'\) on the complexity of \(\problem\).
The usual challenge while performing round elimination is that the description of \(\problem_i\) grows exponentially fast at each iteration, and thus it is fundamental to find \emph{relaxations} of the problem that make \(\problem_i\) easier to solve but hopefully with a simpler description.
Note that while relaxing the description of a problem, one might exaggerate and end up with a problem that is just trivial to solve thwarting all the efforts to understand the complexity of the original problem.
The whole procedure specific to the iterated GHZ problem can be found in \cite{balliu2024quantum}.

We remark that round elimination cannot be used in the quantum world: apart from our result (which separates \qlocal and classical \local), the \emph{no-cloning} principle (that states that quantum states cannot be cloned \cite{d2017quantum}) forbids us any kind of generalization of RE.

\subsection{Networks of \nonsign games}
After we established quantum advantage via \cref{thm:quantum-advantage:iteratedGHZ}, one may wonder whether with more refined games we could achieve even a stronger separation.
In this section we show that, unfortunately, \emph{this is not possible}.

\begin{definition}[Game]\label{def:quantum-advantage:game}
    Let $\Sigma$ be a finite set, and let $m \in \natsPos$ be the number of players.
    We call $\mathfrak{g} \subseteq \Sigma^m \times \Sigma^m$ a \emph{game}.
    Each player $i$ receives one input $x_i \in \Sigma$ and produces one output
    $y_i \in \Sigma$.
    A \emph{move} $\mu = (x,y) \in \Sigma^m \times \Sigma^m$ is valid if $\mu \in
    \mathfrak{g}$.
    We overload the notation so that $\mathfrak{g}(x) = \{ y \in \Sigma^m \mid (x, y) \in \mathfrak{g} \}$.
    We say $\mathfrak{g}$ is \emph{solvable} if, for every $x$, $\mathfrak{g}(x)$
    is non-empty.
  \end{definition}

\paragraph{Non-signaling games.}
Let \(\mathfrak{g}\) be any game of \(m\) players.
We can define the following labeling problem \(\problem\) on any graph \(G= (V,E)\) with \(\abs{V} \ge m\).
In input, a subset \(P \subseteq V\) with \(\abs{P} = m\) is chosen to represent the set of players.
Those nodes receive an input bit and must output another bit so that their joint output solves \(\mathfrak{g}\).
All other nodes can output any bit.
We say that \(\mathfrak{g}\) is \nonsign if there exists a \nonsign outcome that solves \(\problem\) with locality \(0\). 
Clearly, \(\problem\) is not locally checkable, but if we add the further constraint that the diameter of \(G[P]\) is bounded by a constant, then we can make it locally checkable (even if the degree of the graph might be unbounded).

\paragraph{Network of \nonsign games.}
We are given a bipartite graph where the two sets of the bipartition are the set \(W\) of \emph{white nodes} and the set \(B\) of \emph{black nodes}.
White nodes are the real players of the games, while black nodes represent the games that the players are playing.
We assume that each white node has degree \(\maxDeg\) and each black node has degree \(m\).
Each black node represents a \nonsign game that its white neighbors must play.
Also, white nodes can contain arbitrarily complex arithmetic circuits according to which they need to play the games: 
in a sense, the input of a game might depend arbitrarily complex arithmetic operations on the outputs of previous games.
Such a network is called \emph{network of \nonsign games}: it includes networks that one can build with quantum games such as the network in the iterated GHZ problem, and is locally checkable (possibly with unbounded degree).

\cite{balliu2024quantum} proved the following theorem.

\begin{theorem}[\textcite{balliu2024quantum}]\label{thm:quantum-advantage:network-of-non-signaling-games}
    Let \(\maxDeg \in \natsPos\) be a constant.
    For any network of \nonsign games of maximum degree \(\maxDeg\), there exists a classical \local algorithm solving the games in time \(\myO{1}\), where the constant might depend on \(\maxDeg\).  
\end{theorem}

The key ingredient for \cref{thm:quantum-advantage:network-of-non-signaling-games} is that any \nonsign game is completable: for example, for a 2-player game this means that for any Alice's input \(x\) there is some Alice's output \(a\) such that for any
Bob's input \(y\) there is still a valid Bob's output \(b\), and vice versa.
We can exploit completability to solve any network of \nonsign games in a distributed manner: 
Each white
node starts to process its own arithmetic circuit in a topological order. 
As soon as it encounters
a step that involves a game, it sends a message to the black neighbor responsible for that specific
game, together with its own input for that game. The black nodes keep track of the inputs they
have seen so far, and they always pick safe outputs for those players. 
In this way, in two rounds of communication
all white nodes can learn their own output for the game that appears first in their own circuit.
We can then repeat this for each game in a sequential order---thanks to completability, while black nodes will be
always able to assign valid outputs also for players that join the game late. 
The running
time of this algorithm is proportional to the size of the circuit held by a single white node:
for a fixed
LCL (with a finite set of possible local circuits) it will be bounded by some constant.
With this kind of arguments all we can hope for is a separation that is a function \(f(\maxDeg)\) of the degree of the graph, but we still do not have an LCL problem (in the strict sense) separating \local and \qlocal by a function \(f(n)\) of the number of nodes of the graph \(n\).
We conclude this brief survey with the major open question in the field.

\begin{question}[Open]
    Is there any LCL problem that \qlocal can solve asymptotically faster (as a function of the number of nodes) than classical \local?
\end{question} 
\printbibliography 
@inproceedings{akbari_et_al:LIPIcs.ICALP.2023.10,
  author = {Amirreza Akbari and Navid Eslami and Henrik Lievonen and Darya Melnyk and Joona S{\"a}rkij{\"a}rvi and Jukka Suomela},
  editor = {Kousha Etessami and Uriel Feige and Gabriele Puppis},
  title = {Locality in Online, Dynamic, Sequential, and Distributed Graph Algorithms},
  booktitle = {50th International Colloquium on Automata, Languages, and Programming, {ICALP} 2023, July 10-14, 2023, Paderborn, Germany},
  series = {LIPIcs},
  volume = {261},
  pages = {10:1--10:20},
  publisher = {Schloss Dagstuhl - Leibniz-Zentrum f{\"{u}}r Informatik},
  year = {2023},
  doi = {10.4230/LIPICS.ICALP.2023.10}
}

@article{naor1995,
  author = {Moni Naor and Larry J. Stockmeyer},
  title = {What Can be Computed Locally?},
  journal = {SIAM Journal on Computing},
  volume = {24},
  number = {6},
  pages = {1259--1277},
  year = {1995},
  doi = {10.1137/S0097539793254571}
}

@inproceedings{naor1993,
  author       = {Moni Naor and
                  Larry J. Stockmeyer},
  editor       = {S. Rao Kosaraju and
                  David S. Johnson and
                  Alok Aggarwal},
  title        = {What can be computed locally?},
  booktitle    = {Proceedings of the Twenty-Fifth Annual {ACM} Symposium on Theory of
                  Computing, May 16-18, 1993, San Diego, CA, {USA}},
  pages        = {184--193},
  publisher    = {{ACM}},
  year         = {1993},
  doi          = {10.1145/167088.167149}
}

@article{arfaoui2014,
  author = {Heger Arfaoui and Pierre Fraigniaud},
  title = {What can be computed without communications?},
  journal = {SIGACT News},
  volume = {45},
  number = {3},
  pages = {82--104},
  year = {2014},
  doi = {10.1145/2670418.2670440}
}

@inproceedings{gavoille2009,
  author = {Cyril Gavoille and Adrian Kosowski and Marcin Markiewicz},
  editor = {Idit Keidar},
  title = {What Can Be Observed Locally?},
  booktitle = {Distributed Computing, 23rd International Symposium, {DISC} 2009, Elche, Spain, September 23-25, 2009. Proceedings},
  series = {Lecture Notes in Computer Science},
  volume = {5805},
  pages = {243--257},
  publisher = {Springer},
  year = {2009},
  doi = {10.1007/978-3-642-04355-0_26}
}

@article{gavoille2009coloring,
  author       = {Cyril Gavoille and
                  Ralf Klasing and
                  Adrian Kosowski and
                  Lukasz Kuszner and
                  Alfredo Navarra},
  title        = {On the complexity of distributed graph coloring with local minimality
                  constraints},
  journal      = {Networks},
  volume       = {54},
  number       = {1},
  pages        = {12--19},
  year         = {2009},
  doi          = {10.1002/NET.20293}
}

@inproceedings{gavoille2007,
  author       = {Cyril Gavoille and
                  Ralf Klasing and
                  Adrian Kosowski and
                  Alfredo Navarra},
  editor       = {Andrzej Pelc},
  title        = {On the Complexity of Distributed Greedy Coloring},
  booktitle    = {Distributed Computing, 21st International Symposium, {DISC} 2007,
                  Lemesos, Cyprus, September 24-26, 2007, Proceedings},
  series       = {Lecture Notes in Computer Science},
  volume       = {4731},
  pages        = {482--484},
  publisher    = {Springer},
  year         = {2007},
  doi          = {10.1007/978-3-540-75142-7\_37}
}

@inproceedings{coiteuxroy2023,
  author       = {Xavier Coiteux{-}Roy and
                  Francesco D'Amore and
                  Rishikesh Gajjala and
                  Fabian Kuhn and
                  Fran{\c{c}}ois Le Gall and
                  Henrik Lievonen and
                  Augusto Modanese and
                  Marc{-}Olivier Renou and
                  Gustav Schmid and
                  Jukka Suomela},
  editor       = {Bojan Mohar and
                  Igor Shinkar and
                  Ryan O'Donnell},
  title        = {No Distributed Quantum Advantage for Approximate Graph Coloring},
  booktitle    = {Proceedings of the 56th Annual {ACM} Symposium on Theory of Computing,
                  {STOC} 2024, Vancouver, BC, Canada, June 24-28, 2024},
  pages        = {1901--1910},
  publisher    = {{ACM}},
  year         = {2024},
  doi          = {10.1145/3618260.3649679}
}

@article{chang19hierarchy,
  author = {Yi-Jun Chang and Seth Pettie},
  title = {A Time Hierarchy Theorem for the {LOCAL} Model},
  journal = {SIAM Journal on Computing},
  volume = {48},
  number = {1},
  pages = {33--69},
  year = {2019},
  doi = {10.1137/17M1157957}
}

@inproceedings{ghaffari2018derandomizing,
  author = {Mohsen Ghaffari and David G. Harris and Fabian Kuhn},
  editor = {Mikkel Thorup},
  title = {On Derandomizing Local Distributed Algorithms},
  booktitle = {59th {IEEE} Annual Symposium on Foundations of Computer Science, {FOCS} 2018, Paris, France, October 7-9, 2018},
  pages = {662--673},
  publisher = {{IEEE} Computer Society},
  year = {2018},
  doi = {10.1109/FOCS.2018.00069}
}

@article{holroyd2016,
  author = {Alexander E. Holroyd and Thomas M. Liggett},
  title = {{Finitely Dependent Coloring}},
  journal = {Forum of Mathematics, Pi},
  eid = {e9},
  volume = {4},
  year = {2016},
  doi = {10.1017/fmp.2016.7}
}

@article{holroyd2018,
  author = {Alexander E. Holroyd and Tom Hutchcroft and Avi Levy},
  title = {{Finitely dependent cycle coloring}},
  journal = {Electronic Communications in Probability},
  volume = {23},
  year = {2018},
  doi = {10.1214/18-ecp118}
}

@inproceedings{fischer_ghaffari2017sublogarithmic,
  author = {Manuela Fischer and Mohsen Ghaffari},
  editor = {Andr{\'e}a W. Richa},
  title = {Sublogarithmic Distributed Algorithms for {L}ov{\'a}sz Local Lemma, and the Complexity Hierarchy},
  booktitle = {31st International Symposium on Distributed Computing, {DISC} 2017, October 16-20, 2017, Vienna, Austria},
  series = {LIPIcs},
  volume = {91},
  pages = {18:1--18:16},
  publisher = {Schloss Dagstuhl - Leibniz-Zentrum f{\"{u}}r Informatik},
  year = {2017},
  doi = {10.4230/LIPICS.DISC.2017.18}
}

@article{chang_kopelowitz_pettie2019exp_separation,
  author = {Yi-Jun Chang and Tsvi Kopelowitz and Seth Pettie},
  title = {An Exponential Separation between Randomized and Deterministic Complexity in the {LOCAL} Model},
  journal = {SIAM Journal on Computing},
  volume = {48},
  number = {1},
  pages = {122--143},
  year = {2019},
  doi = {10.1137/17M1117537}
}

@inproceedings{ghaffari2017,
  author = {Mohsen Ghaffari and Fabian Kuhn and Yannic Maus},
  editor = {Hamed Hatami and Pierre McKenzie and Valerie King},
  title = {On the complexity of local distributed graph problems},
  booktitle = {Proceedings of the 49th Annual {ACM} {SIGACT} Symposium on Theory of Computing, {STOC} 2017, Montreal, QC, Canada, June 19-23, 2017},
  pages = {784--797},
  publisher = {{ACM}},
  year = {2017},
  doi = {10.1145/3055399.3055471}
}

@inproceedings{linial87,
  author = {Nathan Linial},
  title = {Distributive Graph Algorithms-Global Solutions from Local Data},
  booktitle = {28th Annual Symposium on Foundations of Computer Science, Los Angeles, California, USA, 27-29 October 1987},
  pages = {331--335},
  publisher = {{IEEE} Computer Society},
  year = {1987},
  doi = {10.1109/SFCS.1987.20}
}

@article{linial92,
  author = {Nathan Linial},
  title = {Locality in Distributed Graph Algorithms},
  journal = {SIAM Journal on Computing},
  volume = {21},
  number = {1},
  pages = {193--201},
  year = {1992},
  doi = {10.1137/0221015}
}

@inproceedings{balliu18lcl-complexity,
  author = {Alkida Balliu and Juho Hirvonen and Janne H. Korhonen and Tuomo Lempi{\"a}inen and Dennis Olivetti and Jukka Suomela},
  editor = {Ilias Diakonikolas and David Kempe and Monika Henzinger},
  title = {New classes of distributed time complexity},
  booktitle = {Proceedings of the 50th Annual {ACM} {SIGACT} Symposium on Theory of Computing, {STOC} 2018, Los Angeles, CA, USA, June 25-29, 2018},
  pages = {1307--1318},
  publisher = {{ACM}},
  year = {2018},
  doi = {10.1145/3188745.3188860}
}

@inproceedings{balliu19lcl-decidability,
  author = {Alkida Balliu and Sebastian Brandt and Yi-Jun Chang and Dennis Olivetti and Mika{\"e}l Rabie and Jukka Suomela},
  editor = {Peter Robinson and Faith Ellen},
  title = {The Distributed Complexity of Locally Checkable Problems on Paths is Decidable},
  booktitle = {Proceedings of the 2019 {ACM} Symposium on Principles of Distributed Computing, {PODC} 2019, Toronto, ON, Canada, July 29 - August 2, 2019},
  pages = {262--271},
  publisher = {{ACM}},
  year = {2019},
  doi = {10.1145/3293611.3331606}
}

@inproceedings{balliu19mm,
  author = {Alkida Balliu and Sebastian Brandt and Juho Hirvonen and Dennis Olivetti and Mika{\"e}l Rabie and Jukka Suomela},
  editor = {David Zuckerman},
  title = {Lower Bounds for Maximal Matchings and Maximal Independent Sets},
  booktitle = {60th {IEEE} Annual Symposium on Foundations of Computer Science, {FOCS} 2019, Baltimore, Maryland, USA, November 9-12, 2019},
  pages = {481--497},
  publisher = {{IEEE} Computer Society},
  year = {2019},
  doi = {10.1109/FOCS.2019.00037}
}

@article{balliu20almost-global,
  author = {Alkida Balliu and Sebastian Brandt and Dennis Olivetti and Jukka Suomela},
  title = {Almost global problems in the {LOCAL} model},
  journal = {Distributed Computing},
  volume = {34},
  number = {4},
  pages = {259--281},
  year = {2021},
  doi = {10.1007/S00446-020-00375-2}
}

@inproceedings{balliu20lcl-randomness,
  author = {Alkida Balliu and Sebastian Brandt and Dennis Olivetti and Jukka Suomela},
  editor = {Yuval Emek and Christian Cachin},
  title = {How much does randomness help with locally checkable problems?},
  booktitle = {{PODC} '20: {ACM} Symposium on Principles of Distributed Computing, Virtual Event, Italy, August 3-7, 2020},
  pages = {299--308},
  publisher = {{ACM}},
  year = {2020},
  doi = {10.1145/3382734.3405715}
}

@inproceedings{balliu21lcl-congest,
  author = {Alkida Balliu and Keren Censor-Hillel and Yannic Maus and Dennis Olivetti and Jukka Suomela},
  editor = {Seth Gilbert},
  title = {Locally Checkable Labelings with Small Messages},
  booktitle = {35th International Symposium on Distributed Computing, {DISC} 2021, October 4-8, 2021, Freiburg, Germany (Virtual Conference)},
  series = {LIPIcs},
  volume = {209},
  pages = {8:1--8:18},
  publisher = {Schloss Dagstuhl - Leibniz-Zentrum f{\"{u}}r Informatik},
  year = {2021},
  doi = {10.4230/LIPICS.DISC.2021.8}
}

@inproceedings{balliu22mending,
  author = {Alkida Balliu and Juho Hirvonen and Darya Melnyk and Dennis Olivetti and Joel Rybicki and Jukka Suomela},
  editor = {Merav Parter},
  title = {Local Mending},
  booktitle = {Structural Information and Communication Complexity - 29th International Colloquium, {SIROCCO} 2022, Paderborn, Germany, June 27-29, 2022, Proceedings},
  series = {Lecture Notes in Computer Science},
  volume = {13298},
  pages = {1--20},
  publisher = {Springer},
  year = {2022},
  doi = {10.1007/978-3-031-09993-9_1}
}

@inproceedings{balliu22regular-trees,
  author = {Alkida Balliu and Sebastian Brandt and Yi-Jun Chang and Dennis Olivetti and Jan Studen{\'y} and Jukka Suomela},
  editor = {Christian Scheideler},
  title = {Efficient Classification of Locally Checkable Problems in Regular Trees},
  booktitle = {36th International Symposium on Distributed Computing, {DISC} 2022, October 25-27, 2022, Augusta, Georgia, {USA}},
  series = {LIPIcs},
  volume = {246},
  pages = {8:1--8:19},
  publisher = {Schloss Dagstuhl - Leibniz-Zentrum f{\"{u}}r Informatik},
  year = {2022},
  doi = {10.4230/LIPICS.DISC.2022.8}
}

@article{balliu22rooted-trees,
  author = {Alkida Balliu and Sebastian Brandt and Yi-Jun Chang and Dennis Olivetti and Jan Studen{\'y} and Jukka Suomela and Aleksandr Tereshchenko},
  title = {Locally checkable problems in rooted trees},
  journal = {Distributed Computing},
  volume = {36},
  number = {3},
  pages = {277--311},
  year = {2023},
  doi = {10.1007/S00446-022-00435-9}
}

@inproceedings{balliu22so-simple,
  author = {Alkida Balliu and Janne H. Korhonen and Fabian Kuhn and Henrik Lievonen and Dennis Olivetti and Shreyas Pai and Ami Paz and Joel Rybicki and Stefan Schmid and Jan Studen{\'y} and Jukka Suomela and Jara Uitto},
  editor = {Telikepalli Kavitha and Kurt Mehlhorn},
  title = {Sinkless Orientation Made Simple},
  booktitle = {2023 Symposium on Simplicity in Algorithms, {SOSA} 2023, Florence, Italy, January 23-25, 2023},
  pages = {175--191},
  publisher = {{SIAM}},
  year = {2023},
  doi = {10.1137/1.9781611977585.CH17}
}

@inproceedings{brandt16lll,
  author = {Sebastian Brandt and Orr Fischer and Juho Hirvonen and Barbara Keller and Tuomo Lempi{\"a}inen and Joel Rybicki and Jukka Suomela and Jara Uitto},
  editor = {Daniel Wichs and Yishay Mansour},
  title = {A lower bound for the distributed {L}ov{\'a}sz local lemma},
  booktitle = {Proceedings of the 48th Annual {ACM} {SIGACT} Symposium on Theory of Computing, {STOC} 2016, Cambridge, MA, USA, June 18-21, 2016},
  pages = {479--488},
  publisher = {{ACM}},
  year = {2016},
  doi = {10.1145/2897518.2897570}
}

@inproceedings{brandt17grid-lcl,
  author = {Sebastian Brandt and Juho Hirvonen and Janne H. Korhonen and Tuomo Lempi{\"a}inen and Patric R. J. {\"O}sterg{\r{a}}rd and Christopher Purcell and Joel Rybicki and Jukka Suomela and Przemyslaw Uznanski},
  editor = {Elad Michael Schiller and Alexander A. Schwarzmann},
  title = {{LCL} Problems on Grids},
  booktitle = {Proceedings of the {ACM} Symposium on Principles of Distributed Computing, {PODC} 2017, Washington, DC, USA, July 25-27, 2017},
  pages = {101--110},
  publisher = {{ACM}},
  year = {2017},
  doi = {10.1145/3087801.3087833}
}

@inproceedings{Brandt2019automatic,
  author = {Sebastian Brandt},
  editor = {Peter Robinson and Faith Ellen},
  title = {An Automatic Speedup Theorem for Distributed Problems},
  booktitle = {Proceedings of the 2019 {ACM} Symposium on Principles of Distributed Computing, {PODC} 2019, Toronto, ON, Canada, July 29 - August 2, 2019},
  pages = {379--388},
  publisher = {{ACM}},
  year = {2019},
  doi = {10.1145/3293611.3331611}
}

@inproceedings{chang16exponential,
  author = {Yi-Jun Chang and Tsvi Kopelowitz and Seth Pettie},
  editor = {Irit Dinur},
  title = {An Exponential Separation between Randomized and Deterministic Complexity in the {LOCAL} Model},
  booktitle = {{IEEE} 57th Annual Symposium on Foundations of Computer Science, {FOCS} 2016, 9-11 October 2016, Hyatt Regency, New Brunswick, New Jersey, {USA}},
  pages = {615--624},
  publisher = {{IEEE} Computer Society},
  year = {2016},
  doi = {10.1109/FOCS.2016.72}
}

@article{hirvonen14local-maxcut,
  author = {Juho Hirvonen and Joel Rybicki and Stefan Schmid and Jukka Suomela},
  title = {Large Cuts with Local Algorithms on Triangle-Free Graphs},
  journal = {The Electronic Journal of Combinatorics},
  volume = {24},
  number = {4},
  pages = {4},
  year = {2017},
  doi = {10.37236/6862}
}

@inproceedings{kuhn2004,
  author = {Fabian Kuhn and Thomas Moscibroda and Roger Wattenhofer},
  editor = {Soma Chaudhuri and Shay Kutten},
  title = {What cannot be computed locally!},
  booktitle = {Proceedings of the Twenty-Third Annual {ACM} Symposium on Principles of Distributed Computing, {PODC} 2004, St. John's, Newfoundland, Canada, July 25-28, 2004},
  pages = {300--309},
  publisher = {{ACM}},
  year = {2004},
  doi = {10.1145/1011767.1011811}
}

@inproceedings{legall2019,
  author = {Fran{\c{c}}ois {Le Gall} and Harumichi Nishimura and Ansis Rosmanis},
  editor = {Rolf Niedermeier and Christophe Paul},
  title = {Quantum Advantage for the {LOCAL} Model in Distributed Computing},
  booktitle = {36th International Symposium on Theoretical Aspects of Computer Science, {STACS} 2019, March 13-16, 2019, Berlin, Germany},
  series = {LIPIcs},
  volume = {126},
  pages = {49:1--49:14},
  publisher = {Schloss Dagstuhl - Leibniz-Zentrum f{\"{u}}r Informatik},
  year = {2019},
  doi = {10.4230/LIPICS.STACS.2019.49}
}

@article{korman-2011-global-knowledge,
  author = {Amos Korman and Jean-S{\'e}bastien Sereni and Laurent Viennot},
  title = {Toward more localized local algorithms: removing assumptions concerning global knowledge},
  journal = {Distributed Computing},
  volume = {26},
  number = {5-6},
  pages = {289--308},
  year = {2013},
  doi = {10.1007/S00446-012-0174-8}
}

@inproceedings{legall2018,
  author = {Fran{\c{c}}ois {Le Gall} and Fr{\'e}d{\'e}ric Magniez},
  title = {Sublinear-Time Quantum Computation of the Diameter in {CONGEST} Networks},
  booktitle = {Proceedings of the 2018 ACM Symposium on Principles of Distributed Computing},
  publisher = {ACM},
  pages = {337--346},
  year = {2018},
  doi = {10.1145/3212734.3212744}
}

@article{magniez2022,
  author = {Fr{\'e}d{\'e}ric Magniez and Ashwin Nayak},
  title = {Quantum Distributed Complexity of Set Disjointness on a Line},
  journal = {ACM Transactions on Computation Theory},
  volume = {14},
  number = {1},
  pages = {5:1--5:22},
  year = {2022},
  doi = {10.1145/3512751}
}

@inproceedings{wu2022,
  author = {Xudong Wu and Penghui Yao},
  editor = {Alessia Milani and Philipp Woelfel},
  title = {Quantum Complexity of Weighted Diameter and Radius in {CONGEST} Networks},
  booktitle = {{PODC} '22: {ACM} Symposium on Principles of Distributed Computing, Salerno, Italy, July 25 - 29, 2022},
  pages = {120--130},
  publisher = {{ACM}},
  year = {2022},
  doi = {10.1145/3519270.3538441}
}

@article{wang2021,
author = {Wang, Changsheng and Wu, Xudong and Yao, Penghui},
title = {Complexity of Eccentricities and All-Pairs Shortest Paths in the Quantum CONGEST Model},
journal = {SPIN},
volume = {11},
number = {03},
pages = {2140007},
year = {2021},
doi = {10.1142/S2010324721400075},
eprint = {https://doi.org/10.1142/S2010324721400075}
}

@inproceedings{izumi2019,
  author = {Taisuke Izumi and Fran{\c{c}}ois {Le Gall}},
  editor = {Peter Robinson and Faith Ellen},
  title = {Quantum Distributed Algorithm for the All-Pairs Shortest Path Problem in the {CONGEST-CLIQUE} Model},
  booktitle = {Proceedings of the 2019 {ACM} Symposium on Principles of Distributed Computing, {PODC} 2019, Toronto, ON, Canada, July 29 - August 2, 2019},
  pages = {84--93},
  publisher = {{ACM}},
  year = {2019},
  doi = {10.1145/3293611.3331628}
}

@inproceedings{censorhillel2022,
  author = {Keren Censor-Hillel and Orr Fischer and Fran{\c{c}}ois {Le Gall} and Dean Leitersdorf and Rotem Oshman},
  editor = {Mark Braverman},
  title = {Quantum Distributed Algorithms for Detection of Cliques},
  booktitle = {13th Innovations in Theoretical Computer Science Conference, {ITCS} 2022, January 31 - February 3, 2022, Berkeley, CA, {USA}},
  series = {LIPIcs},
  volume = {215},
  pages = {35:1--35:25},
  publisher = {Schloss Dagstuhl - Leibniz-Zentrum f{\"{u}}r Informatik},
  year = {2022},
  doi = {10.4230/LIPICS.ITCS.2022.35}
}

@inproceedings{izumi2020,
  author = {Taisuke Izumi and Fran{\c{c}}ois {Le Gall} and Fr{\'e}d{\'e}ric Magniez},
  editor = {Christophe Paul and Markus Bl{\"a}ser},
  title = {Quantum Distributed Algorithm for Triangle Finding in the {CONGEST} Model},
  booktitle = {37th International Symposium on Theoretical Aspects of Computer Science, {STACS} 2020, March 10-13, 2020, Montpellier, France},
  series = {LIPIcs},
  volume = {154},
  pages = {23:1--23:13},
  publisher = {Schloss Dagstuhl - Leibniz-Zentrum f{\"{u}}r Informatik},
  year = {2020},
  doi = {10.4230/LIPICS.STACS.2020.23}
}

@inproceedings{apeldoorn2022,
  author = {Joran van Apeldoorn and Tijn de Vos},
  editor = {Alessia Milani and Philipp Woelfel},
  title = {A Framework for Distributed Quantum Queries in the {CONGEST} Model},
  booktitle = {{PODC} '22: {ACM} Symposium on Principles of Distributed Computing, Salerno, Italy, July 25 - 29, 2022},
  pages = {109--119},
  publisher = {{ACM}},
  year = {2022},
  doi = {10.1145/3519270.3538413}
}

@misc{Olivetti2019,
  author = {Dennis Olivetti},
  title = {Round Eliminator: a tool for automatic speedup simulation},
  url = {https://github.com/olidennis/round-eliminator},
  year = {2019}
}

@article{naor1991,
  author = {Moni Naor},
  title = {A Lower Bound on Probabilistic Algorithms for Distributive Ring Coloring},
  journal = {SIAM Journal on Discrete Mathematics},
  volume = {4},
  number = {3},
  pages = {409--412},
  year = {1991},
  doi = {10.1137/0404036}
}

@book{d2017quantum,
  author = {Giacomo Mauro D'Ariano and Giulio Chiribella and Paolo Perinotti},
  title = {Quantum Theory from First Principles: An Informational Approach},
  publisher = {Cambridge University Press},
  year = {2016},
  doi = {10.1017/9781107338340}
}

@inproceedings{fraigniaud2024,
  author = {Pierre Fraigniaud and Ma{\"e}l Luce and Fr{\'e}d{\'e}ric Magniez and Ioan Todinca},
  editor = {Ran Gelles and Dennis Olivetti and Petr Kuznetsov},
  title = {Even-Cycle Detection in the Randomized and Quantum {CONGEST} Model},
  booktitle = {Proceedings of the 43rd {ACM} Symposium on Principles of Distributed Computing, {PODC} 2024, Nantes, France, June 17-21, 2024},
  pages = {209--219},
  publisher = {{ACM}},
  year = {2024},
  doi = {10.1145/3662158.3662767}
}

@article{balliu2024shared-randomness,
  author       = {Alkida Balliu and
                  Mohsen Ghaffari and
                  Fabian Kuhn and
                  Augusto Modanese and
                  Dennis Olivetti and
                  Mika{\"{e}}l Rabie and
                  Jukka Suomela and
                  Jara Uitto},
  title        = {Shared Randomness Helps with Local Distributed Problems},
  journal      = {CoRR},
  volume       = {abs/2407.05445},
  year         = {2024},
  doi          = {10.48550/ARXIV.2407.05445}
}

@article{aaronson2022,
  author       = {Scott Aaronson},
  title        = {How Much Structure Is Needed for Huge Quantum Speedups?},
  journal      = {CoRR},
  volume       = {abs/2209.06930},
  year         = {2022},
  doi          = {10.48550/ARXIV.2209.06930}
  }

@inproceedings{derbel2008,
  author       = {Bilel Derbel and
                  Cyril Gavoille and
                  David Peleg and
                  Laurent Viennot},
  editor       = {Rida A. Bazzi and
                  Boaz Patt{-}Shamir},
  title        = {On the locality of distributed sparse spanner construction},
  booktitle    = {Proceedings of the Twenty-Seventh Annual {ACM} Symposium on Principles
                  of Distributed Computing, {PODC} 2008, Toronto, Canada, August 18-21,
                  2008},
  pages        = {273--282},
  publisher    = {{ACM}},
  year         = {2008},
  doi          = {10.1145/1400751.1400788}
}

@inproceedings{elkin2007,
  author       = {Michael Elkin},
  editor       = {Indranil Gupta and
                  Roger Wattenhofer},
  title        = {A near-optimal distributed fully dynamic algorithm for maintaining
                  sparse spanners},
  booktitle    = {Proceedings of the Twenty-Sixth Annual {ACM} Symposium on Principles
                  of Distributed Computing, {PODC} 2007, Portland, Oregon, USA, August
                  12-15, 2007},
  pages        = {185--194},
  publisher    = {{ACM}},
  year         = {2007},
  doi          = {10.1145/1281100.1281128}
}

@article{bogdanov2013,
	author = {Ilya I. Bogdanov},
	copyright = {arXiv.org perpetual, non-exclusive license},
	doi = {10.48550/arXiv.1311.2844},
	eprint = {1311.2844},
	eprinttype = {arXiv},
	journal = {CoRR},
	publisher = {arXiv},
	title = {Examples of topologically highly chromatic graphs with locally small chromatic number},
	doi = {10.48550/arXiv.1311.2844},
	volume = {abs/1311.2844},
	year = {2013}
}

@article{mohar2002,
  author       = {Bojan Mohar and
                  Paul D. Seymour},
  title        = {Coloring Locally Bipartite Graphs on Surfaces},
  journal      = {J. Comb. Theory {B}},
  volume       = {84},
  number       = {2},
  pages        = {301--310},
  year         = {2002},
  doi          = {10.1006/JCTB.2001.2086}
}

@article{mohar2013,
	author = {Bojan Mohar and G{\'a}bor Simonyi and G{\'a}bor Tardos},
	doi = {10.1007/S00493-013-2771-Y},
	journal = {Combinatorica},
	number = {4},
	pages = {467--495},
	title = {Local chromatic number of quadrangulations of surfaces},
	volume = {33},
	year = {2013}
}

@inproceedings{akbari2024,
  author       = {Amirreza Akbari and
                  Xavier Coiteux{-}Roy and
                  Francesco D'Amore and
                  Fran{\c{c}}ois Le Gall and
                  Henrik Lievonen and
                  Darya Melnyk and
                  Augusto Modanese and
                  Shreyas Pai and
                  Marc{-}Olivier Renou and
                  V{\'{a}}clav Rozhon and
                  Jukka Suomela},
  title        = {Online Locality Meets Distributed Quantum Computing},
  arxiv = {2403.01903},
  booktitle = {Proceedings of the 57th Annual {ACM} Symposium on Theory of Computing, {STOC} 2025, Prague, CZ, Czech Republic, June 23-27, 2025},
  publisher = {{ACM}},
  year = {2025}
}

@inproceedings{dhar24rand,
  author       = {Anubhav Dhar and
                  Eli Kujawa and
                  Henrik Lievonen and
                  Augusto Modanese and
                  Mikail Muftuoglu and
                  Jan Studen{\'{y}} and
                  Jukka Suomela},
  editor       = {Silvia Bonomi and
                  Letterio Galletta and
                  Etienne Rivi{\`{e}}re and
                  Valerio Schiavoni},
  title        = {Local Problems in Trees Across a Wide Range of Distributed Models},
  booktitle    = {28th International Conference on Principles of Distributed Systems,
                  {OPODIS} 2024, December 11-13, 2024, Lucca, Italy},
  series       = {LIPIcs},
  volume       = {324},
  pages        = {27:1--27:17},
  publisher    = {Schloss Dagstuhl - Leibniz-Zentrum f{\"{u}}r Informatik},
  year         = {2024},
  doi          = {10.4230/LIPICS.OPODIS.2024.27}
}

@inproceedings{chang20,
  author       = {Yi{-}Jun Chang},
  editor       = {Hagit Attiya},
  title        = {The Complexity Landscape of Distributed Locally Checkable Problems
                  on Trees},
  booktitle    = {34th International Symposium on Distributed Computing, {DISC} 2020,
                  October 12-16, 2020, Virtual Conference},
  series       = {LIPIcs},
  volume       = {179},
  pages        = {18:1--18:17},
  publisher    = {Schloss Dagstuhl - Leibniz-Zentrum f{\"{u}}r Informatik},
  year         = {2020},
  doi = {10.4230/LIPIcs.DISC.2020.18}
}

@inproceedings{grunau2022,
  author       = {Christoph Grunau and
                  V{\'{a}}clav Rozhon and
                  Sebastian Brandt},
  editor       = {Alessia Milani and
                  Philipp Woelfel},
  title        = {The Landscape of Distributed Complexities on Trees and Beyond},
  booktitle    = {{PODC} '22: {ACM} Symposium on Principles of Distributed Computing,
                  Salerno, Italy, July 25 - 29, 2022},
  pages        = {37--47},
  publisher    = {{ACM}},
  year         = {2022},
  doi = {10.1145/3519270.3538452}
}

@article{balliu2024quantum,
  author       = {Alkida Balliu and
                  Sebastian Brandt and
                  Xavier Coiteux{-}Roy and
                  Francesco D'Amore and
                  Massimo Equi and
                  Fran{\c{c}}ois Le Gall and
                  Henrik Lievonen and
                  Augusto Modanese and
                  Dennis Olivetti and
                  Marc{-}Olivier Renou and
                  Jukka Suomela and
                  Lucas Tendick and
                  Isadora Veeren},
  title        = {Distributed Quantum Advantage for Local Problems},
  arxiv = {2411.03240},
  booktitle = {Proceedings of the 57th Annual {ACM} Symposium on Theory of Computing, {STOC} 2025, Prague, CZ, Czech Republic, June 23-27, 2025},
  publisher = {{ACM}},
  year = {2025}
}

@article{Brassard2004QuantumP,
  title={Quantum Pseudo-Telepathy},
  author={Gilles Brassard and Anne Broadbent and Alain Tapp},
  journal={Foundations of Physics},
  year={2005},
  volume={35},
  pages={1877-1907},
  doi = {10.1007/s10701-005-7353-4}
}

@article{BennettS98,
  author       = {Charles H. Bennett and
                  Peter W. Shor},
  title        = {Quantum Information Theory},
  journal      = {{IEEE} Trans. Inf. Theory},
  volume       = {44},
  number       = {6},
  pages        = {2724--2742},
  year         = {1998},
  doi          = {10.1109/18.720553}
}
\end{document}